\documentclass[11pt,urlcolor=blue, linkcolor=blue]{article} 
\usepackage{cite}

\usepackage{amsmath, amsthm, amssymb,slashed}
\usepackage{mathtools}

\usepackage{ifpdf}
\ifpdf
  \usepackage[pdftex]{graphicx}
  \usepackage{epstopdf}
\else
  \usepackage[dvips]{graphicx}
\fi
\usepackage[letterpaper]{geometry}
\parskip=\baselineskip

\usepackage{tablefootnote}

\usepackage[usenames, dvipsnames]{xcolor}

\allowdisplaybreaks[1]
\numberwithin{equation}{section}

\newcommand{\cblue}[1]{\textcolor{blue}{#1}}

\usepackage{upgreek}

\usepackage{tabularx}

\DeclareMathAlphabet{\mathpzc}{OT1}{pzc}{m}{it}

\newcommand{\bea}{\begin{eqnarray}}
\newcommand{\eea}{\end{eqnarray}}
\def\be{\begin{equation}}
\def\ee{\end{equation}}

\newcommand{\re}{\hspace{1pt}\mathrm{e}}

\usepackage{color}
\usepackage[colorlinks,citecolor=blue]{hyperref}
\definecolor{red}{rgb}{1,0,0}
\definecolor{blue}{rgb}{0,0,1}
\definecolor{dblue}{rgb}{0,0,0.4}
\definecolor{green}{rgb}{0,1,0}
\definecolor{black}{rgb}{0,0,0}
\definecolor{white}{rgb}{1,1,1}

\definecolor{brn}{rgb}{.8,.4,.0}
\definecolor{redo}{rgb}{1,.5,.0}
\definecolor{ddgrn}{rgb}{0,0.4,0}
\definecolor{dgrn}{rgb}{0,0.55,0}
\definecolor{dbl}{rgb}{0,0,0.5}

\usepackage[bbgreekl]{mathbbol}
\usepackage{amscd}

\newcommand{\ii}{\hspace{1pt}\mathrm{i}\hspace{1pt}}
\newcommand{\dd}{\hspace{1pt}\mathrm{d}}

\newcommand{\Refe}[1]{Ref.~\cite{#1}}
\newcommand{\Eq}[1]{Eq.~(\ref{#1})} 
 
\newcommand{\eqn}[1]{Eq.~(\ref{#1})}

\newcommand{\Tr}{{\rm Tr}} 
 
\renewcommand{\Im}{{\rm Im}} 
\renewcommand{\Re}{{\rm Re}}

\newcommand{\prt}{\partial}

\newcommand{\bpm}{\begin{pmatrix}}
\newcommand{\epm}{\end{pmatrix}}
\newcommand{\bmm}{\begin{matrix}}
\newcommand{\emm}{\end{matrix}}

\newcommand{\cD}{ {\cal D} }

\def\CQ{{\cal Q}}

\newcommand{\Z}{\mathbb{Z}}
\newcommand{\bC}{\mathbb{C}}
\newcommand{\R}{\mathbb{R}}

\def\Z{{\mathbb{Z}}}

\def\Tr{{\mathrm{Tr}}}

\def\bZ{{\mathbf{Z}}}

\DeclareRobustCommand\sWang%{{\includegraphics[height=4.25ex]{1280px-Wang}}}
{\cblue{\includegraphics[height=3.6ex]{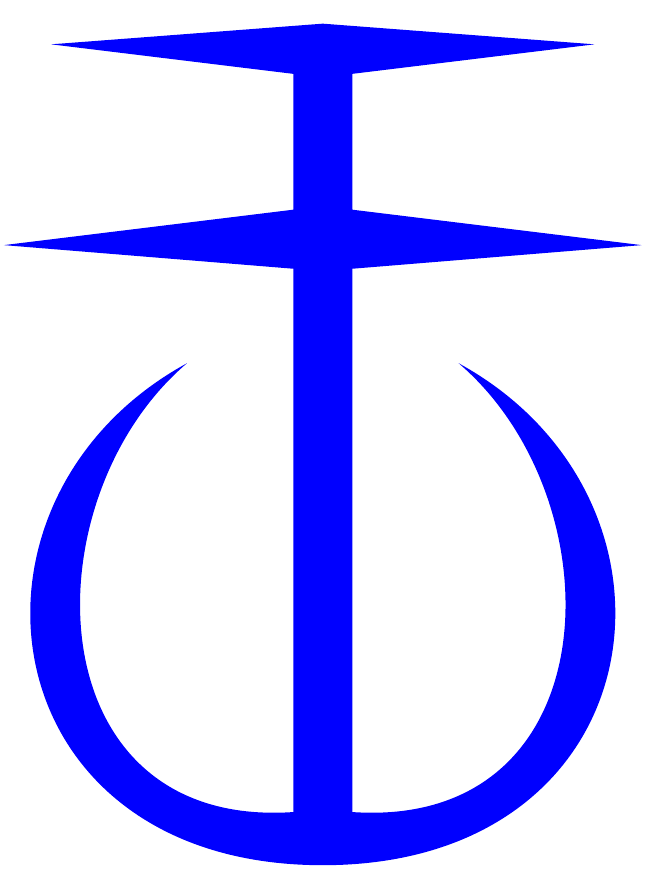}}}

\newcommand{\Wangfootnote}[1]{%
\let\oldthefootnote=\thefootnote%
\stepcounter{mpfootnote}%
\addtocounter{footnote}{-1}%
\renewcommand{\thefootnote}{\sWang}
\footnote{#1}%
\let\thefootnote=\oldthefootnote%
}

\DeclareRobustCommand\sXu%
{\cblue{\includegraphics[height=5.2ex]{Xu-Images-Xu.pdf}}}

\newcommand{\Xufootnote}[1]{%
\let\oldthefootnote=\thefootnote%
\stepcounter{mpfootnote}%
\addtocounter{footnote}{-1}%
\renewcommand{\thefootnote}{\sXu}
\footnote{#1}%
\let\thefootnote=\oldthefootnote%
}

\DeclareRobustCommand\sYau%
{\cblue{\includegraphics[height=3.6ex]{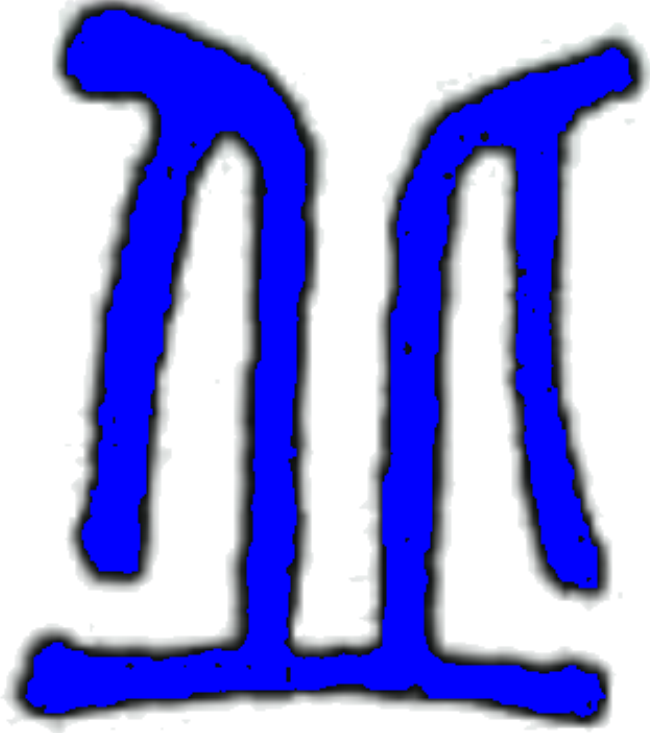}}}

\newcommand{\Yaufootnote}[1]{%
\let\oldthefootnote=\thefootnote%
\stepcounter{mpfootnote}%
\addtocounter{footnote}{-1}%
\renewcommand{\thefootnote}{\sYau}
\footnote{#1}%
\let\thefootnote=\oldthefootnote%
}

\usepackage{enumitem,cleveref}% http://ctan.org/pkg/{enumitem,cleveref}
\newcommand{\nn}{\nonumber}

\usepackage{comment}
\usepackage{verbatim}

\def \- {\!\smallsetminus\!}

\newcommand{\tO}{{\text O}}

\newcommand{\SO}{{\rm SO}}

\newcommand{\U}{{\rm U}}

\newcommand{\rN}{{\rm N}}

\def \Im{\operatorname{Im}}
\def \rH{\operatorname{H}}

\def \Z{\mathbb{Z}}

\newcommand{\Sec}[1]{Sec.~\ref{#1}}

\usepackage{tikz}
\usetikzlibrary{calc}

\usepackage[all,cmtip]{xy}
\usepackage{tikz-cd}
\usetikzlibrary{matrix}
\usetikzlibrary{knots}

\usetikzlibrary{calc}

\usepackage[all,cmtip]{xy}
\usepackage{tikz-cd}
\usepackage{datetime}

\newenvironment{myfont}[2][]{\csname#2\endcsname[#1]}{}

\usepackage{makecell} % For thicker hlines in Table

\usepackage{slashed}
\usepackage[makeroom]{cancel}
\usepackage[normalem]{ulem} %for \sout
\usepackage{soul}

\usepackage[X2,T1]{fontenc}
\DeclareSymbolFont{cyrillic}{X2}{cmr}{m}{n}
\SetSymbolFont{cyrillic}{bold}{X2}{cmr}{bx}{n}
\DeclareMathSymbol{\khk}{\mathord}{cyrillic}{139}
\DeclareMathSymbol{\zh}{\mathord}{cyrillic}{117}
\DeclareMathSymbol{\zhen}{\mathord}{cyrillic}{182}
\DeclareMathSymbol{\zhe}{\mathord}{cyrillic}{134}

\def\re{{\mathrm{e}}}

\def \U{\mathrm{U}}
\def \rO{\mathrm{O}}

\newcommand{\tm}{\hspace{1pt}\mathrm{m}}

\newcommand{\Wplusfootnote}[1]{%
\let\oldthefootnote=\thefootnote%
\stepcounter{mpfootnote}%
\addtocounter{footnote}{-1}%
\renewcommand{\thefootnote}{{W$^+$}}
\footnote{#1}%
\let\thefootnote=\oldthefootnote%
}

\newcommand{\Yfootnote}[1]{%
\let\oldthefootnote=\thefootnote%
\stepcounter{mpfootnote}%
\addtocounter{footnote}{-1}%
\renewcommand{\thefootnote}{{Y$^+$}}
\footnote{#1}%
\let\thefootnote=\oldthefootnote%
}

\begin{document}
\begin{titlepage}
\begin{flushright}
%cond-mat/hep/math/
\end{flushright}
\vskip 12.mm
%\vskip .50in

\begin{center}

%Symmetric
{\bf\LARGE{%Quantum 
%%Higher-Spin %/%
%Higher-Rank  %Abelian 
%Tensor  Non-Commutative Gauge
%Field Theory:  \\[5.5mm]  
%%Fully 
% %Non-Abelian 
% Gauged Fractonic Matter versus New Sigma Models, 
%\\[5.5mm]  
%Superfluids and Non-Abelian Vortices 
%\\[6.8mm] 
Non-Abelian 
Gauged Fracton Matter Field Theory: \\[5.5mm]  
New Sigma Models, 
Superfluids and %Non-Abelian 
Vortices 
\\[6.8mm] 
}}

%\vskip-0.mm 
\vskip.5cm
\quad\quad\quad
\Large{
Juven Wang
$^{1,2}${\Wangfootnote{e-mail: {\tt jw@cmsa.fas.harvard.edu
} (Corresponding Author)  \href{http://sns.ias.edu/~juven/}{http://sns.ias.edu/$\sim$juven/}}} 
% \flushright   \quad\quad\quad\quad\quad\quad\quad\quad\quad\quad\quad\quad \quad\quad 
%\quad    Kai Xu$^{3}$, {\Xufootnote{e-mail: {\tt  kaixu@math.harvard.edu}}}
and  \; Shing-Tung Yau
$^{1, 3,4}${\Yaufootnote{e-mail: {\tt  yau@math.harvard.edu}
\hfill  December 2019 \\[2mm]
%\flushleft
{
%\begin{center}
%\centering
 %\quad\quad
  \emph{Dedicated to %an upcoming 
  the
70 year anniversary of Ginzburg-Landau %superconductivity
theory} [Zh. Ekaper. Teoret. Fiz. 20, 1064 (1950)] 
 \emph{in 2020}.
%\end{center}
\flushleft .
}
 %\quad\quad\quad\quad \quad\quad  
} 
}
}
%\quad\quad \quad

\vskip.5cm
{\small{\textit{$^1${Center of Mathematical Sciences and Applications, Harvard University,  Cambridge, MA 02138, USA} \\}}
}
\vskip.2cm
{\small{\textit{$^2$School of Natural Sciences, Institute for Advanced Study,  Einstein Drive, Princeton, NJ 08540, USA}\\}}
\vskip.2cm
{\small{\textit{$^3$ Department of Mathematics, Harvard University, Cambridge, MA 02138, USA}\\}}
\vskip.2cm
{\small{\textit{$^4$ Department of Physics, Harvard University, Cambridge, MA 02138, USA}\\}}

\end{center}
\vskip.35cm
\baselineskip 12pt
\begin{abstract}

By gauging a higher-moment polynomial degree %-(m-1) 
global symmetry and a discrete charge conjugation (i.e., particle-hole) symmetry 
coupled to matter fields (two symmetries mutually non-commutative),
we derive a new 
class of higher-rank 
tensor non-abelian gauge field theory with dynamically gauged fractonic matter fields: 
Non-abelian gauged fractons interact with
a hybrid class of higher-rank (symmetric or generic non-symmetric) tensor gauge theory and anti-symmetric tensor topological field theory, 
generalizing [arXiv:1909.13879, 1911.01804]. 
We also apply a
quantum phase transition similar to that between insulator v.s. superfluid/superconductivity 
(U(1) symmetry disordered phase described by a topological gauge theory or a disordered Sigma model  
v.s.~U(1) global/gauge symmetry-breaking ordered phase described by a Sigma model with a U(1) target space underlying Goldstone modes):
We can regard our tensor gauge theories as disordered phases, and we transient to their new ordered phases
by deriving new Sigma models in 
continuum field theories.
While one low energy theory is captured by degrees of freedom of rotor or scalar modes,
another side of low energy theory
has vortices and superfluids --- 
we  
explore non-abelian vortices (two types of vortices mutually interacting non-commutatively beyond an ordinary group structure) and their Cauchy-Riemann relation.
%\\[4mm]
 
%due to a non-commutativity between the polynomial and  charge conjugation symmetry, 
%(but non-group structure) 

%\today, \currenttime

\end{abstract}
\end{titlepage}
%(e.g. conformal) 
%Lorentzian 
%in 1865 
%in 1954. 
%our theory 
%giving a first hint
%``gauge group-analogous structure''

%Secs
% 1 

\renewcommand{\eqref}{\eqn}

  \pagenumbering{arabic}
    \setcounter{page}{2}
    
\tableofcontents   

%\newpage

%%%%%%%%%%%%%%%%%%%%%%%%%%%%%%%%%%%%%%%%
%%%%%%%%%%%%%%%%%%%%%%%%%%%%%%%%%%%%%%%%
%%%%%%%%%%%%%%%%%%%%%%%%%%%%%%%%%%%%%%%%
%%%%%%%%%%%%%%%%%%%%%%%%%%%%%%%%%%%%%%%%
%%%%%%%%%%%%%%%%%%%%%%%%%%%%%%%%%%%%%%%%

\noindent

\section{Introduction %and Summary
and Overview of Previous Works}
\label{sec:intro}

Fracton order (see a recent review \cite{RahulNandkishore2018sel1803.11196} in condensed matter)
concerns new conservation laws imposed on the energetic excitations (such that the particle excitations are known as fractons)
of quantum systems which have significant restrictions on their mobility:
\begin{enumerate}
\item 
Excitations cannot move without creating additional excitations (commonly known as fractons), 
\item
Excitations can only move in certain subdimensional or subsystem directions (for 0-dimensional excitations known as subdimensional particles).
\end{enumerate}
The origins of such constraints are new conservation laws from conserved quantities of higher-moments, including dipole moments \cite{Pretko2018jbi1807.11479} 
(relevant for a vector global symmetry in field theory \cite{SeibergF2019, Wang2019aiq1909.13879}), quadrupole moments, or generalized
multipole moment (relevant for the so-called the polynomial global symmetry \cite{Gromov2018nbv1812.05104, Wang2019aiq1909.13879, WXY2-1910-1911.01804} 
or the polynomial shift symmetries \cite{GriffinPetrHorava2014bta1412.1046} in field theory), etc.
%%%%%%%
The composite excitation of each mobility-restricted excitations are
however mobile.
The mobility constraint of fracton phases is also related to quantum glassy dynamics \cite{2005PRL0404182Chamon, 1108.2051CastelnovoChamon}.
%%%%%%%
Follow the previous work of \Refe{Wang2019aiq1909.13879, WXY2-1910-1911.01804},
motivated by the fracton order in condensed matter \cite{RahulNandkishore2018sel1803.11196},
we continue extending and developing this framework by including the dynamically gauged matter fields in the higher-rank tensor gauge 
theory in a $d+1$ dimensional spacetime (e.g. $d+1$d, over a flat spacetime manifold ${M^{d+1}}$, and we should focus on Cartesian coordinates
${\R^{d+1}}$).\footnote{
In this article, we follow the notations and definitions given in \Refe{Wang2019aiq1909.13879, WXY2-1910-1911.01804}.
We attempt to be succinct in this article, 
thus we will directly guide the readers to refer the sections/materials derived in \Refe{Wang2019aiq1909.13879} and \Refe{WXY2-1910-1911.01804}.
}
The important new ingredient in our present work is that the gauge structure can be non-commutative (i.e., the so-called non-abelian), while still \emph{coupling to the matter fields} ---
thus a partial goal of our present work is to derive a new non-abelian tensor gauged fracton field theory (that has gauge interactions also \emph{coupling with gauged matter fields}).

To recall, the field-theoretic models given in \Refe{Wang2019aiq1909.13879, WXY2-1910-1911.01804}
offer some unified features that we may summarize via examples that also connect to the literature:
\begin{enumerate}[leftmargin=2.0mm, label=\textcolor{blue}{\arabic*}., ref=\textcolor{blue}{\arabic*}]
\item 
\emph{An ungauged matter field theory of higher-moment global symmetry without gauge fields} 
(e.g., an ungauged abelian theory with a degree-0 ordinary symmetry that encodes 
Schr\"odinger  \cite{1926PhRvSchrodinger} or Klein-Gordon type field theory \cite{1926Klein, 1926Gordon}
(see \Sec{sec:Degree-0-Klein-Gordon}),
with a degree-1 polynomial symmetry pioneered in Pretko's work \cite{Pretko2018jbi1807.11479} (see \Sec{sec:Degree-1-Pretko's}) and its
general higher-moment degree-$(\tm-1)$ polynomial generalization \cite{Gromov2018nbv1812.05104, WXY2-1910-1911.01804} (see \Sec{sec:Degree-m-1})):
For example, there is a field theory captured by the Lagrangian term with a covariant derivate term
$( P_{i_1,\cdots,i_{\tm}}(\Phi,\cdots,\prt^{\tm}\Phi) )$ or $P_J$ given in Sec.~2.1 and Sec.~3.1 of \Refe{WXY2-1910-1911.01804}.
The schematic path integrals $\bZ$ are: 
\bea
\bZ&=& \int[\cD \Phi][\cD \Phi^\dagger]\exp(\ii \int_{M^{d+1}} \dd^{d+1}x \Big( P_{i_1,\cdots,i_{\tm}}(\Phi,\cdots,\prt^{\tm}\Phi) \Big)
             \Big( P^{i_1,\cdots,i_{\tm}}(\Phi^\dagger,\cdots,\prt^{\tm}\Phi^\dagger) \Big) ). 
\eea
The dynamical complex scalar fields 
\bea
\Phi :=\Phi(x) =\Phi(\vec{x},t)  \text{ and }  \Phi^\dagger=\Phi^\dagger(x)=\Phi^\dagger (\vec{x},t) \in \bC
\eea 
are summed over in a schematic path integral.
The Lagrangian term and $\bZ$ are invariant under the global symmetry transformation.

\item
\emph{A pure abelian or non-abelian 
higher-rank tensor gauge theory} (\emph{without coupling to gauged matter field}): 

The abelian case is widely studied in various works in condensed matter literature:
\Refe{2016arXiv160108235RRasmussenYouXu, Pretko2016kxt1604.05329, Pretko2016lgv1606.08857, Pretko2017xar1707.03838, Slagle2018kqf1807.00827}.

But a non-abelian version of fractonic theory is not much explored.
Recent progress on non-abelian fracton orders from \Refe{PremHuangSong2018jsn1806.04687, BulmashMaissamBarkeshli2019taq1905.05771, PremWilliamson2019etl1905.06309} are mostly built from lattice models with a discrete gauge group (or a discrete gauge structure in general).

We will take an alternative route to non-abelian fracton via the field theory.
A rank-2 non-abelian higher-rank tensor gauge theory with a continuous gauge structure is proposed firstly in \Refe{Wang2019aiq1909.13879}.
The most general form of rank-m non-abelian 
higher-rank tensor gauge theory is given by a schematic path integral in  \Refe{WXY2-1910-1911.01804}'s Sec.~2.1
\begin{multline} \label{eq:U(1)poly-Z2C-gauge}
\bZ_{\underset{\text{asym-BF}}{\text{rk-m-sym-$A$}}}: =
\int
(\prod_{I=1}^{N}[\cD A_{I,{i_1,\cdots,i_{\tm}}}] [\cD B_I]  [\cD C_I])\\
%[\cD A_{i_1,\cdots,i_{\tm}} ][\cD B][\cD C]
\exp(\ii \int_{M^{d+1}} \dd^{d+1}x \Big( 
\sum_{I=1}^N  |\hat F^c_{I,\mu,\nu,i_2,\cdots,i_{\tm}}|^2  + \frac{ 2}{2 \pi}  \sum_{I=1}^{N} B_I \dd C_I\Big)) \cdot \omega_{d+1}(\{C_I\}).
\end{multline} 
The cocycle
$\omega_{d+1} \in \rH^{d+1}((\Z_2^C)^N, \R/\Z)$ is a group cohomology data \cite{DijkgraafWitten1989pz1990} that we can take the continuum 
topological quantum field theory (TQFT) formulation 
of discrete gauge theory (see References therein \cite{1602.05951WWY, Putrov2016qdo1612.09298, Wang2018edf1801.05416, Wang1901.11537WWY} and the overview \cite{Wang2019aiq1909.13879}).
The $I$ is an index for specifying the different copies/layers of tensor gauge theories, the
cocycle $\omega_{d+1}$ couples different copies/layers of tensor gauge theories together.  
Thus the cocycle $\omega_{d+1}$  gives rise to the interlayer interaction effects.
The index $I$ may be neglected for simplicity below.
%%%%%%%%%
 The real-valued abelian gauge field strength $F_{\mu,\nu,i_2,\cdots,i_{\tm}} \in \R$ is promoted into a new 
 complex-valued non-abelian gauge field strength 
$\hat F^c_{\mu,\nu,i_2,\cdots,i_{\tm}} \in \bC$
after gauging a discrete charge conjugation $\Z_2^C$ (i.e., particle-hole) 
symmetry \cite{Wang2019aiq1909.13879, WXY2-1910-1911.01804}: 
\bea
{\hat F^c_{\mu,\nu,i_2,\cdots,i_{\tm}}  :=D_\mu^c A_{\nu,i_2,\cdots,i_{\tm}}  -D_{\nu }^c A_{\mu,i_2,\cdots,i_{\tm}}
:=(\prt_\mu - \ii g_c C_\mu) A_{\nu,i_2,\cdots,i_{\tm}} -(\prt_{\nu} - \ii g_c C_\nu ) A_{\mu,i_2,\cdots,i_{\tm}} },
\eea
while
$| \hat F^c_{\mu,\nu,i_2,\cdots,i_{\tm}}|^2 :=  \hat F^c_{\mu,\nu,i_2,\cdots,i_{\tm}}  \hat F^{\dagger c\; {\mu,\nu,i_2,\cdots,i_{\tm}}}$.
Here are the field contents:
\begin{itemize}
%%%%%%%%%%%%%%%%%%%%%%%%%%%%%%%%%%%%%%%%
\item 
The $A \in \R$ can be chosen to be a fully-symmetric rank-m real-valued tensor gauge field. 
\item 
The $B  \in \R$ is a $(d-1)$-th $\Z_2$-cohomology class in terms of $\Z_2$-discrete gauge theory, or in the continuum formulated as a 
$(d-1)$-form (an anti-symmetric rank-$(d-1)$ tensor) real-valued gauge field. 
The $B$ plays the role of a Lagrangian multiplier to set $C$ to be flat.
\item 
The $C  \in \R$ is a $\Z_2$-cohomology class  in terms of $\Z_2$-discrete gauge theory, or in the continuum formulated as 
a $1$-form (a rank-$1$ tensor) real-valued gauge field.
\end{itemize}
%%%%%%%%%%%%%%%%%%%%%%%%%%%%%%%%%%%%%%%%
\item
\emph{An abelian gauge theory coupling to gauged matter field}: This is pioneered in Pretko's \cite{Pretko2018jbi1807.11479} for the rank-2 tensor fields,
while we can use the most general form for the rank-m tensor gauge field $A=A_{i_1,i_2,\cdots,i_{\tm}}$ given in  \Refe{WXY2-1910-1911.01804} Sec.~2.1's and Sec.~3.1's  schematic path integral:
\be \hspace{-12mm}
\label{eq:U(1)vector-gauge-matter-rank-m}
{\bZ_{\text{rk-2-sym-$\Phi$}} =\int[\cD A][\cD \Phi][\cD \Phi^\dagger]\exp(\ii \int_{M^{d+1}} \dd^{d+1}x \Big( %\frac{1}{g^2}
 |F_{\mu,\nu,i_2,\cdots,i_{\tm}}|^2 +
|D_{}^A[\{\Phi \}]|^2 +V(|\Phi|^2)
\Big))}.
\ee
$$
|D_{}^A[\{\Phi\}]|^2  := |R_{}|^2 :=  (R_{}) (R_{}^{\dagger}) =( P_{} - \ii g A_{} \CQ_{}) ( P_{}^{\dagger} + \ii g A_{} \CQ_{}^{\dagger}).
$$
The $D_{}^A[\{\Phi \}]  :=R_{} \equiv P_{} - \ii g A_{} \CQ_{}$ is defined in Sec.~3.1 of \Refe{WXY2-1910-1911.01804}.
Here $P_{}$ and $\CQ_{}$ are polynomials of $\Phi_{}$ and its differential of $\prt^{\ell} \Phi_{}$ for some power of ${\ell}$.
Here $P_{}$ and $\CQ_{}$ are uniquely determined by the polynomial $Q_{}(x)$ in the higher-moment global symmetry %$\Phi_I \to \re^{\ii Q_I(x)}\Phi_I$.
\bea
\Phi_I\to\re^{\ii Q_I(x)}\Phi_I := \re^{\ii \big(\Lambda_{I;i_1,\dots,i_{{\tm}-1}}  x_{i_1} \dots x_{i_{\tm}-1} +\dots+ \Lambda_{I; i,j}  x_ix_j + \Lambda_{I; i}  x_i + \Lambda_{I;  0}\big)}\Phi_I,
\eea
shown in \Refe{WXY2-1910-1911.01804}.
We denote such a polynomial symmetry
as ${{\rm{U}}(1)_{\rm{poly}}}$ following \cite{WXY2-1910-1911.01804}, see a review in \Sec{sec:degree-(m-1)-polynomial-symmetry}
%%%%%%%%%%%%%%%%%%%%%%%%%%%%%%%%%%%%%%%%

\end{enumerate}

What else topics have not yet been done in the literature but should be formulated?
We will focus on these two open issues:
%This issue is one of the two focuses of our present work.
%
\begin{enumerate}[leftmargin=2.0mm, label=\textcolor{blue}{\arabic*}., ref=\textcolor{blue}{\arabic*}]
\item
\emph{A non-abelian gauge theory coupling to gauged matter field}: 

Previous works only did the abelian gauged matter theory, or the non-abelian gauge theory without coupling to (fractonic) matter fields  \cite{Wang2019aiq1909.13879, WXY2-1910-1911.01804}.
In Sec \ref{sec:Non-Abelian-Gauged-Fractonic-Matter-Theories},
we provide a systematic framework for non-abelian gauged fractonic matter field theories.

In order to facilitate such a non-abelian gauged matter formulation,
we sometimes trade a single complex component 
$
\Phi =  \Phi_{\Re} + \ii \Phi_{\Im}\in \bC
$
into two real components
$\begin{pmatrix} \Phi_{\Re} \in \R \\ \Phi_{\Im} \in \R \end{pmatrix}$.
 For the U(1) polynomial symmetry viewpoint, the $\Phi \in \bC$ is more natural.
The $\Z_2^C$ (particle-hole or particle-anti-particle symmetry) transformation acts on $\Phi \to \Phi^\dagger$ in the complex
U(1) basis, but the $\Z_2^C$ acts on the 2-component field naturally as:
$$
\begin{pmatrix} \Phi_{\Re} \\ \Phi_{\Im} \end{pmatrix}
 \to \begin{pmatrix} \Phi_{\Re} \\- \Phi_{\Im} \end{pmatrix}.
 $$
 To introduce the non-abelian gauge coupling to the matter fields,
 we will need to introduce several new types of
 gauge derivatives.\footnote{Note that some of such gauge derivatives 
 are \emph{not gauge covariant} under the gauge transformations, due to the non-Gaussian nature and higher-moment terms appear already in the 
 fractonic matter field theories. But we will be able to construct the new types of 
 gauge covariant terms and gauge-invariant Lagrangian in \Sec{sec:Non-Abelian-Gauged-Fractonic-Matter-Theories}.
 }
 For example, even for the simplest degree-0 polynomial symmetry with a rank-1 tensor gauge field (1-form gauge field $A_{\mu}$), we require: 
\bea
D_{\mu}^{c,\Im} \Phi
&:=&
\prt_{\mu} \Phi -\ii g_c C_\mu \Phi_{\Im},\\
D_{\mu}^A \Phi
&:=&
(\prt_{\mu} - \ii g A_{\mu})\Phi
=
(\prt_\mu \Phi_\Re + g A_\mu\Phi_{\Im})
+\ii (\prt_\mu \Phi_{\Im} - g A_\mu\Phi_\Re), \\
D_{\mu}^{A,c,\Im} \Phi
&:=&
D_{\mu}^A \Phi -\ii g_c C_\mu
\Phi_{\Im}=(\prt_{\mu} - \ii g A_{\mu})\Phi -\ii g_c C_\mu \Phi_{\Im}. \label{eq:Gauge-covariant-1-derivative-nab} \\
D_{\mu}^c &:=&(\prt_\mu - \ii g_c C_{\mu,I}).\\
D_\mu^c A_{\nu,i_2,\cdots,i_{\tm}}
&:=&(\prt_\mu - \ii g_c C_\mu) A_{\nu,i_2,\cdots,i_{\tm}} .
\eea
 The first line $D_{\mu}^{c,\Im}$ has the $C$ gauge field that only couples to the charged matter $\Phi_{\Im}$ (the imaginary component) under  $\Z_2^C$.
 The second line  $D_{\mu}^A$ is the gauge covariant derivative of 1-form gauge field $A_\mu$ after gauging the ordinary 0-form U(1) symmetry (the degree-0 polynomial symmetry).
 The third line $D_{\mu}^{A,c,\Im} \Phi$ shows the gauge derivative on $\Phi$ involving both the $\U(1) \rtimes \Z_2^C = \rm{O}(2)$ gauge fields.
 The forth and the fith line shows that for the fields charged under $\Z_2^C$ (e.g. the rank-m symmetric tensor $ A_{\nu,i_2,\cdots,i_{\tm}}
 \to  -A_{\nu,i_2,\cdots,i_{\tm}}$ is charged under $\Z_2^C$), 
 then the gauge derivative is $D_{\mu}^c$. The $g_c$ is a $\Z_2^C$ gauge coupling denoted explicitly for the convenience. 
 
%%%%%%
We will present explicit examples to gauge both higher-moment and charge conjugation global symmetries
including the matter in \Sec{sec:Non-Abelian-Gauged-Fractonic-Matter-Theories}.
%%%%%%%%%%
%%%%%%%%%%
%%%%%%%%%%
%%%%%%%%%%

\item
\emph{A new type of Sigma model}: 
We formulate a new type of Sigma model that can move between the ordered and disordered phases
of these
higher-rank non-abelian 
tensor  
field theories with  
fully gauged fractonic matter.\footnote{In terms of 
the old Landau-Ginzburg paradigm, this is
related to the Sigma model formulation of Landau-Ginzburg theory \cite{Ginzburg1950srGinzburgLandau}.
%\cred{Landau-Ginzburg theory and Sigma model} 
}
Similar to the familiar quantum phase transition between insulator v.s. superfluid/superconductivity 
 \cite{{Fisher-Lee},{Dasgupta-Halperin},{Nelson}}
(U(1) symmetry disorder described by a topological gauge theory or a disordered Sigma model  
v.s.~U(1) global/gauge symmetry-breaking order described by a Sigma model with a U(1) target space with Goldstone modes),
%(U(1) symmetry-breaking order described by a Sigma model with a U(1) target space with Goldstone modes) 
%(U(1) symmetry disorder described by a disordered Sigma model or a topological gauge theory),
we can regard our tensor gauge theory as a disordered phase, and we drive to its new ordered phase
by deriving a new Sigma model in terms of continuum field theory.

Very recently, the superfluid and vortices of an abelian version of pure fractonic theories (without gauge fields) are studied in \cite{Yuan2019gehPengYe1911.02876}.
Two new ingredients in our work, which are not present in \Refe{Yuan2019gehPengYe1911.02876},
are the facts that we include the gauge field interactions (thus we include the additional long-range entanglements) and we also include the
non-abelian gauge-matter interactions.

\end{enumerate}

%%%%%%%%%%%%%%%%%%%%%%%%%%%%%%%%%%%%%%%%
%%%%%%%%%%%%%%%%%%%%%%%%%%%%%%%%%%%%%%%%
%%%%%%%%%%%%%%%%%%%%%%%%%%%%%%%%%%%%%%%%
%%%%%%%%%%%%%%%%%%%%%%%%%%%%%%%%%%%%%%%%
%%%%%%%%%%%%%%%%%%%%%%%%%%%%%%%%%%%%%%%%
%%%%%%%%%%%%%%%%%%%%%%%%%%%%%%%%%%%%%%%%

\section{From Abelian Gauge Fractonic Theories to Non-Abelian Gauged Fractonic Matter Theories}

\label{sec:Non-Abelian-Gauged-Fractonic-Matter-Theories}

\subsection{Degree-0 polynomial symmetry to Schr\"odinger  or Klein-Gordon type field theory}
\label{sec:Degree-0-Klein-Gordon}

\subsubsection{Global-covariant 1-derivative} 

Suppose we want to construct a field theory that preserves a {degree-0 polynomial} symmetry with a polynomial $Q(x)=\Lambda_0$.
Then a complex scalar field transforms as
$\Phi :=\Phi(x) =\Phi(\vec{x},t)    \in \bC $
\bea
\Phi\to  \re^{\ii Q(x)}\Phi= \re^{\ii  \Lambda_0}\Phi,
\eea
while its log transforms as
\bea
\log \Phi \to  \log\Phi+ {\ii  \Lambda_0}.
\eea
Take the derivative $\prt_{x_i}:= \prt_{i}$ respect to coordinates on both sides, we can eliminate $\prt_{i} \Lambda_0$ thus we get an invariant term:
\bea
\prt_{i} \log\Phi   \to \prt_{i}\log\Phi,
\eea
This means $\prt_{i} \log\Phi$ is invariant under the global symmetry transformation.  
We can also define 
\bea
\prt_{i} \log\Phi :=
		\frac{P_{i}(\Phi,\prt\Phi)}{\Phi}= \frac{{\prt_{i} \Phi }}{\Phi}
\eea
Under $\Phi\to  \re^{\ii Q(x)}\Phi$, since the $\prt_{i} \log\Phi$ is invariant and the denominator $\Phi \to \re^{\ii Q(x)}\Phi$ is covariant, 
so does the numerator ${P_{i}(\Phi,\prt\Phi)} = {\prt_{i} \Phi } \to \re^{\ii Q(x)} {\prt_{i} \Phi }$
 is covariant.
%%%%%%%%
Namely, both the numerator ${P_{i}(\Phi,\prt\Phi)}$ and ${\Phi}$ are global-covariant under $\Phi\to  \re^{\ii Q(x)}\Phi$,
in order to maintain the $\prt_{i} \log\Phi$ to be invariant.
Here ${P_{i}(\Phi,\prt\Phi)}$ denotes some functional ${P_{i}}$ that depends on fields $\Phi$ or its derivative $\prt\Phi$.

For convenience, we will call such construction a \emph{global-covariant 1-derivative} 
\bea
{P_{i}(\Phi,\prt\Phi)}:={\prt_{i} \Phi }
\eea to facilitate its further generalization later.
We also construct
 a \emph{globally invariant} 
 Lagrangian 
\bea \label{eq:global-covariant 1-derivative-L}
 |{P_{i}}|^2 + V(|\Phi|^2)
\eea
 that contains a potential term $V(|\Phi|^2)$ and a kinetic term\footnote{The raising and the lowering indices
 are merely used for contractions and summation, e.g. we sum over $i$ indices in ${\prt_{i} \Phi }{\prt^{i} \Phi^\dagger}$.
 The conversion between the raising and lowering indices do not involve spacetime metrics, we only consider the flat spacetime.}
 \bea
 |{P_{i}}|^2 := {P_{i}}(\Phi){P^{i}}(\Phi^\dagger)={\prt_{i} \Phi }{\prt^{i} \Phi^\dagger }. 
 \eea
In this way, based on the systematic method of  \Refe{WXY2-1910-1911.01804}, we can re-derive a Lagrangian formulation of
Schr\"odinger equation in 1925 \cite{1926PhRvSchrodinger} and Klein-Gordon theory in 1926 \cite{1926Klein, 1926Gordon}
for complex scalar fields.

\subsubsection{Gauge-covariant 1-derivative}

To gauge a {degree-0 polynomial} symmetry,	we rewrite $Q(x)$ as a local gauge parameter $\eta(x)$,
\bea
\Phi &\to&  \re^{\ii \eta(x)}\Phi , \label{eq:phi-gauge-eta}\\
\prt_{i} \log\Phi &\to& \prt_{i} \log\Phi  +  \ii \prt_{i}\eta(x).  \label{eq:log-phi-gauge-eta}
\eea
Then $\prt_{i} \log\Phi $ is no longer an invariant term.
This implies that we can write a new gauge-covariant operator $D_{i}^A[\{\Phi \}]$ via
combining $P_{i}$ and $A_{i}$:
	\bea
	P_{i}(\Phi,\prt\Phi) &:=& \prt_{i} \Phi  \to\re^{\ii  \eta(x)} (P_{i}(\Phi,\prt\Phi) +\ii \prt_{i} \eta(x)).\\
        	A_{i}  &\to &A_{i} +\frac{1}{g} \prt_{i}  \eta.\\
	        D_{i}^A[\{\Phi \}] &:=&P_{i }(\Phi,\prt\Phi) - \ii g A_{i} \Phi =  \prt_{i} \Phi - \ii g A_{i} \Phi.
	        \label{eq:D-phi}\\
D_{i}^A[\{\Phi \}] &\to & \re^{\ii  \eta(x)} D_{i}^A[\{\Phi \}] .
	\eea
To obtain a gauge invariant term, we can pair the gauge-covariant operator with its complex conjugation
so to obtain a gauge invariant Lagrangian
\bea \label{eq:gauge-covariant 1-derivative-L}
\left| D_{i}^A[\{\Phi \}] \right|^2 + V(|\Phi|^2)
=(( \prt_{i}  - \ii g A_{i}) \Phi) (( \prt^{i}  + \ii g A^{i}) \Phi^\dagger)
+ V(|\Phi|^2).
\eea

\subsubsection{Gauge-covariant non-abelian rank-2 field strength}

Notice that in the pure matter theory \Eq{eq:global-covariant 1-derivative-L}
without gauge fields, we already have a degree-0 U(1) global symmetry and a $\Z_2^C$ discrete charge conjugation (i.e., particle-hole) symmetry: 
\bea
 \Phi \mapsto   \Phi^\dagger, \quad
 \eea
 which makes \Eq{eq:global-covariant 1-derivative-L} invariant. 
 It is easy to see that the symmetry group structure is a non-abelian group
 \bea
\U(1) \rtimes \Z_2^C =\SO(2) \rtimes \Z_2^C =\rO(2),
 \eea
 which acts on the $\Phi$ non-commutatively: 
 \bea
U_{\Z_2^C} U_{{\rm{U}}(1)_{}}  \Phi \;  &=&U_{\Z_2^C} ( e^{\ii \eta} \Phi)
=e^{\ii \eta} \Phi^\dagger.\nn\\
 U_{{\rm{U}}(1)_{}} U_{\Z_2^C} \Phi \; &=&U_{{\rm{U}}(1)_{}} (\Phi^\dagger)
=e^{-\ii \eta} \Phi^\dagger.
\eea
After we dynamically gauge the U(1) symmetry to obtain \Eq{eq:gauge-covariant 1-derivative-L},
we can still keep $\Z_2^C$ discrete charge conjugation (i.e., particle-hole) symmetry intact which acts on gauge fields as
\bea 
 A_{i }  \mapsto -A_{i }, \quad \eta(x) \mapsto - \eta(x).
\eea
If we fully gauge $\U(1) \rtimes \Z_2^C =\rO(2)$, we get a non-abelian $\rO(2)$ gauge transformations
which acts also on gauge fields non-commutatively: 
\bea
U_{\Z_2^C} U_{{\rm{U}}(1)_{}} A_{j}&=&U_{\Z_2^C} ( A_{j}  +\frac{1}{g} \prt_j  \eta)
=- A_{j } +\frac{1}{g} \prt_j   \eta.\nn\\
 U_{{\rm{U}}(1)_{}} U_{\Z_2^C} A_{j }&=&U_{{\rm{U}}(1)_{}} ( -A_{j })
=- A_{j } -\frac{1}{g} \prt_j  \eta.\nn
\eea
By promoting the global $\Z_2^C$ to a local symmetry,
we introduce a new 1-form $\Z_2^C$-gauge field $C$ coupling to the
0-form symmetry $\Z_2^C$-charged object $A_{i}$ with a new $g_c$ coupling.
The $\Z_2^C$ local gauge transformation is: 
\bea \label{eq:Z2C-gauge}
A_{i}  \to  \re^{\ii \gamma_c(x)} A_{i}, \quad 
C_i \to C_i +\frac{1}{g_c} \prt_i \gamma_c(x).
\eea
Note $A_{i_1} \in \R$ is in real-valued, so a generic $\re^{\ii \gamma_c(x)}$ complexifies the $A_{i}$. 
Thus we restrict gauge transformation to be only $\Z_2^C$-gauged (not $\U(1)^C$-gauged)
 \bea \label{eq:Z2-phase}
 \re^{\ii \gamma_c(x)} := (-1)^{\gamma_c'(x)} \in \{\pm 1\}, \quad \gamma_c \in \pi \Z, \quad \gamma_c' \in \Z, 
 \eea
 so ${\gamma_c'(x)}$ is an integer and $A_{i} \to \pm A_{i}$ stays in real. 
 In the continuum field theory, 
the restrict gauge transformation is done by coupling to a level-2 BF theory (the  $\Z_2^C$-gauge theory) \cite{Wang2019aiq1909.13879, WXY2-1910-1911.01804}.
 Thus ${\gamma_c'(x)}$ jumps between even or odd integers in $\Z$, while the 
 $\Z_2^C$-gauge transformation can be suitably formulated on a lattice. 
 We can also directly express the above on  a triangulable spacetime manifold or a simplicial complex.
 
 \paragraph{Approach 1:  Gauge 0-form $\Z_2^C$-symmetry of 1-form gauge field $A$\\}
 
Follow \Refe{Wang2019aiq1909.13879}, we also define a new covariant derivative with respect to $\Z_2^C$:
\bea \label{eq}
D_i^c :=(\prt_i - \ii g_c C_i).
\eea
We obtain a combined O(2) gauge transformation of $A_{j}$,  
\bea \label{eq:vector-gaugeA-w-matter}
   A_{j }  &\to&\re^{\ii \gamma_c(x)} A_{j } + (\pm)^{\rtimes} \frac{1}{g}(D_j^c ) ( \eta(x))
   :=\left\{\begin{array}{l} 
   V_{\Z_2} V_{\U(1)}A_{j }= \re^{\ii \gamma_c(x)} A_{j } +  \frac{1}{g}D_j^c \eta, \\
   V_{\U(1)}V_{\Z_2}A_{j }= \re^{\ii \gamma_c(x)} (A_{j } +  \frac{1}{g}D_j^c \eta) \end{array}\right. .
\eea
 We define that a new operation 
 $$
 (\pm)^{\rtimes} \in \{+1,-1\}
 $$ via the above \Eq{eq:vector-gaugeA-w-matter}.
Only when we perform a $\Z_2^C$ first and then  $\U(1)_{}$ gauge transformation second, and when 
$\re^{\ii \gamma_c(x)} =-1$ we have 
$(\pm)^{\rtimes}=-1$, otherwise all the other cases $(\pm)^{\rtimes}=+1$.
This factor $(\pm)^{\rtimes}$ also captures the \emph{non-abelian}-ness of the gauge structure.

\Refe{Wang2019aiq1909.13879, WXY2-1910-1911.01804} defines
a rank-2 \emph{non-abelian} field strength as
\bea \label{eq:rank-2-nAb-F}
\hat F^c_{i_1,i_2}  &:=&{D_{i_1}^c A_{i_2}  -D_{i_2 }^c A_{i_1}}
:={(\prt_{i_1} - \ii g_c C_{i_1}) A_{i_2} -(\prt_{i_2 } - \ii g_c C_{i_2 } ) A_{i_1} },
\eea
with the locally flat $\Z_2^C$-gauge field $C$ imposed by the level-2 BF theory 
$\frac{ 2}{2 \pi} \int B \dd C$. It can be shown that
$\hat F^c_{i_1,i_2}$ is gauge-covariant  under the 
O(2) gauge transformation \Eq{eq:vector-gaugeA-w-matter}:
\bea
\hat F^c_{i_1,i_2}  &\to &\re^{\ii \gamma_c(x)} \hat F^c_{i_1,i_2}. 
\eea
The above rank-2 field strength utilizes the viewpoint of
{gauging 0-form $\Z_2^C$-symmetry of 1-form gauge field $A$.} However, because
$\hat F^c_{i_1,i_2}$ is a O(2) field strength, we can write
$\hat F^c_{i_1,i_2}$ in a conventional way as a $2 \times 2$ matrix like Yang-Mills \cite{PhysRev96191YM1954} did.
We will puesue this alternative way in the next paragraph.

\paragraph{Approach 2: Non-abelian O(2) field strength\\}

Earlier from the U(1) symmetry transformation, 
a single complex component 
\bea
\Phi =  \Phi_{\Re} + \ii \Phi_{\Im}\in \bC
\eea is a more natural view.
From the $\rO(2)=\SO(2) \times \Z_2^C$ view, the 2-component real scalar field 
$
( \Phi_{\Re} \in \R, \\ \Phi_{\Im} \in \R )
$
is natural such that the
$U_{{\rm{U}}(1)_{}}$ 
and $U_{\Z_2^C}$ symmetry transforms the 2-component field as:
\bea
U_{{\rm{U}}(1)_{}}&:& 
\begin{pmatrix} \Phi_{\Re} \\ \Phi_{\Im} \end{pmatrix}
 \to
 \begin{pmatrix}
\cos(\theta) & \sin(\theta)\\
-\sin(\theta) & \cos(\theta)
\end{pmatrix}
  \begin{pmatrix} \Phi_{\Re} \\\Phi_{\Im} \end{pmatrix}.
\\
U_{\Z_2^C} &:&
\begin{pmatrix} \Phi_{\Re} \\ \Phi_{\Im} \end{pmatrix}
 \to \begin{pmatrix} \Phi_{\Re} \\- \Phi_{\Im} \end{pmatrix}.
 \eea
 Let $A$ gauge field be the generator of $U_{{\rm{U}}(1)_{}}$, and $C$
gauge field be the generator of  $U_{\Z_2^C}$.
We can write down the non-abelian O(2) gauge field $X$ 
and field strength $\hat{F}_X$
with a Lie algebra generator as:\footnote{Readers may wonder whether the O(2) Lie algebra generator needs to be traceless. There are however two facts:
\begin{enumerate}[leftmargin=2.0mm, label=\textcolor{blue}{(\arabic*)}., ref=\textcolor{blue}{(\arabic*)}]
\item
It is known that Lie algebra
generators of a semi simple Lie algebra must be traceless.
A Lie algebra is semisimple if it is a direct sum of simple Lie algebras, i.e., non-abelian Lie algebras ${\mathfrak {g}}$ 
whose only ideals are {0} and ${\mathfrak {g}}$ itself. 
However, a one-dimensional Lie algebra (which is necessarily abelian) is by definition not considered a simple Lie algebra, 
although such an algebra has no nontrivial ideals. 
Thus, one-dimensional algebras are not allowed as summands in a semisimple Lie algebra.
\item 
The $C$ is a discrete $\Z_2^C$ 1-form gauge field so $\dd C$ is locally flat. Later on we need to impose the condition to show gauge covariance of field strength.
(In general we do not have to impose equations of motion to show gauge invariance, although in the case with the $\Z_2^C$-gauge field $C$, we do require 
its locally flatness for gauge covariance of
$\hat F^c_{i_1,i_2}$ or $\hat{F}_X$.)
\end{enumerate}
}
\bea
&&\hspace{-10mm}
X = \begin{pmatrix}
0 & A\\
-A & g_c C
\end{pmatrix},
\nn\\
&&\hspace{-10mm}
\hat{F}_X=\dd X - \ii  X X=\dd\begin{pmatrix}
0 & A\\
-A & g_c C
\end{pmatrix}
- \ii 
\begin{pmatrix}
0 & A\\
-A &g_c C
\end{pmatrix}
\begin{pmatrix}
0 & A\\
-A &g_c C
\end{pmatrix}
=
\begin{pmatrix}
0 & \dd A\\
-\dd A &g_c \dd C
\end{pmatrix}
- \ii g_c
\begin{pmatrix}
0 & A C\\
AC & 0
\end{pmatrix}\quad\nn \\
&&=
\begin{pmatrix}
0 & \dd A - \ii g_c A C\\
-\dd A - \ii g_c A C& 0
\end{pmatrix}.
\eea
Here we use the fact that $\dd C$ is locally flat.

The Yang-Mills O(2) field strength kinetic term from $\hat{F}_X$ is proportional to:
\bea
&&\hspace{-10mm}
\Tr[\hat{F}_X \wedge \star \hat{F}_X^\dagger]
=
\Tr[\begin{pmatrix}
0 &  \prt_{\mu} A_{\nu} - \ii  g_c  A_{\mu} C_{\nu}\\
- \prt_{\mu} A_{\nu} - \ii g_c  A_{\mu} C_{\nu}&0 
\end{pmatrix} 
\begin{pmatrix}
0 & - \prt^{\mu} A^{\nu} + \ii  g_c  A^{\mu} C^{\nu}\\
 \prt^{\mu} A^{\nu} + \ii  g_c  A^{\mu} C^{\nu}&0 
\end{pmatrix} 
]\dd^4x \nn\\
%&&=
%\Tr[
%\begin{pmatrix}
%( \prt_{\mu} A_{\nu})^2 + g_c^2 ( A_{\mu} C_{\nu})^2 & .\\
%.&( \prt_{\mu} A_{\nu})^2 +  g_c^2 ( A_{\mu} C_{\nu})^2 
%\end{pmatrix} 
%]\dd^4x \nn\\
&&=
2\Big( ( \prt_{\mu} A_{\nu})^2 + g_c ( A_{\mu} C_{\nu})^2 \Big) \dd^4x. 
\eea
In comparison, the
rank-2 non-Abelian field strength $\hat F^c_{\mu \nu }$ defined in  \Refe{WXY2-1910-1911.01804} and
\Eq{eq:rank-2-nAb-F} outputs
\bea
&&{\hat F^{c}_{\mu \nu }} {\hat F^{c \mu \nu \dagger}} =
((\prt_\mu - \ii g_c C_{\mu}) A_{\nu } -(\prt_{\nu} - \ii g_c C_{\nu} ) A_{\mu })
((\prt^\mu + \ii g_c C^{\mu}) A^{\nu } -(\prt^{\nu} + \ii g_c C^{\nu} ) A^{\mu })\nn \\
&&\propto  \prt_\mu A_{\nu }\prt^\mu A^{\nu }
-( g_c)^2 ( A_{\mu} C_{\nu})^2.
\label{eq:rank-2-nAb-F-L}
\eea
Thus two approaches agree on the Yang-Mills Lagrangian
$\Tr[\hat{F}_X \wedge \star \hat{F}_X^\dagger] \sim {\hat F^{c}_{\mu \nu }} {\hat F^{c \mu \nu \dagger}}$,
 up to a normalization constant.

\paragraph{Gauge-covariant non-abelian rank-2 field strength}

Under the O(2) gauge transformation,
$$A_{j }  \to A_{j }'=\re^{\ii \gamma_c(x)} A_{j } + (\pm)^{\rtimes} \frac{1}{g}(D_j^c ) ( \eta(x))  \eta, \quad
 C_j \to  C_{j }'=C_j +\frac{1}{g_c} \prt_j \gamma_c(x),$$
 wee can explicitly check that the non-abelian rank-2 field strength ${\hat F^{c}_{\mu \nu }}$ is gauge covariant:
 \bea
 &&{\hat F^{c}_{\mu \nu }} :={D_{\mu}^c A_{\nu } -D_{\nu }^c A_{\mu  }
:=(\prt_\mu - \ii g_c C_{\mu}) A_{\nu } -(\prt_{\nu} - \ii g_c C_{\nu} ) A_{\mu }}\nn\\
&&\to \re^{\ii \gamma_c}
{\hat F^{c}_{\mu \nu }} + \frac{1}{g} (D_\mu^c D_\nu^c -D_\nu^c D_\mu^c) \eta_v  
=\re^{\ii \gamma_c}
{\hat F^{c}_{\mu \nu }},
 \eea
where we list down the leading order omitting the potentially {higher power of $\eta$ and $\gamma_c$} terms.
Note that: \bea
 &&(D_\mu^c D_\nu^c -D_\nu^c D_\mu^c)=
(\prt_\mu - \ii g_c C_{\mu})(\prt_{\nu} - \ii g_c C_{\nu} ) - (\prt_{\nu} - \ii g_c C_{\nu} )(\prt_\mu - \ii g_c C_{\mu}) \nn\\
&&
=(\prt_\mu \prt_{\nu}  - \prt_{\nu}  \prt_\mu ) - \ii g_c (\prt_\mu C_{\nu} - \prt_{\nu} C_{\mu})
- \ii g_c ( C_{\nu} \prt_\mu -C_{\mu}  \prt_{\nu} )
- \ii g_c (  C_{\mu}  \prt_{\nu} - C_{\nu} \prt_\mu)
- g_c^2(  C_{\mu}  C_{\nu} 
-C_{\nu}  C_{\mu}) \nn\\
&&=- \ii g_c (\prt_\mu C_{\nu} - \prt_{\nu} C_{\mu})
=- \ii g_c ( \dd C)_{\mu \nu}=0,
\label{eq:DD-DD}
\eea	
where we need to impose the locally flat condition for the $\Z_2^C$ gauge field in 
the last equality.\footnote{See also a related discussion but with a more explicit calculation
on the gauge covariance of field strength  in \Refe{Wang2019aiq1909.13879}}
Thus the gauge covariance is true since we show $(D_\mu^c D_\nu^c -D_\nu^c D_\mu^c)=0$.

\subsubsection{Non-abelian O(2) gauged matter: Polynomial invariant v.s. Yang-Mills method}
\label{sec:degree-0-nab-gauged-M}
Previous work  \cite{Wang2019aiq1909.13879, WXY2-1910-1911.01804} does not couple to non-abelian gauge fields to matter field.
In this work, we propose a systematic method to generate non-abelian gauged matter theories.

 \paragraph{Approach 1:  Polynomial invariant method --- Covariant derivative on the log as an invariant}

For any give complex field $\mathfrak{N} \in \bC$,
such that $\mathfrak{N} = \mathfrak{N}_{\Re} + \ii  \mathfrak{N}_{\Im}$, where the imaginary $ \mathfrak{N}_{\Im} \to - \mathfrak{N}_{\Im}$ 
is charged under $\Z_2^C$-symmetry, thus we define a new derivative 
 \bea
D^{c,\Im}_{\mu} \mathfrak{N} &:=& \prt_{\mu} \mathfrak{N}_{\Re} + \ii  D^{c}_{\mu}  \mathfrak{N}_{\Im}
\equiv\prt_{\mu} \mathfrak{N}_{\Re} + \ii  (\prt_{\mu}- \ii g_c C_{\mu})  \mathfrak{N}_{\Im}.
\eea
For example, for the complex scalar field $\Phi= \Phi_{\Re} + \ii  \Phi_{\Im} \in \bC$, with the real component $\Phi_{\Re} \in \R$ and imaginary  
component $\Phi_{\Im} \in \R$, 
\bea
D^{c,\Im}_{\mu} \Phi &:=& \prt_{\mu} \Phi_{\Re} + \ii  D^{c}_{\mu}  \Phi_{\Im}
\equiv\prt_{\mu} \Phi_{\Re} + \ii  (\prt_{\mu}- \ii g_c C_{\mu})  \Phi_{\Im}.
\eea
Follow \Eq{eq:phi-gauge-eta} and \Eq{eq:log-phi-gauge-eta},
in order to find an O(2) gauge covariant derivative,
we aim to firstly design an invariant term under gauge transformations.
First, we see that $D_\mu^{c,\Im}  \log\Phi$ is not 
invariant under a U(1) part of O(2) gauge transformation $\Phi \to  \re^{\ii \eta(x)}\Phi $,
\bea
&&{D_\mu^{c,\Im}  \log\Phi \to
	D_\mu^{c,\Im}   \log\Phi  +  D_\mu^{c,\Im}(\ii   \eta(x))
	= \frac{ D_\mu^{c,\Im}\Phi +  D_\mu^{c,\Im}(\ii   \eta(x)) \Phi}{\Phi} 
	=\frac{ D_\mu^{c,\Im}\Phi + \ii (D_\mu^c  \eta) \Phi}{\Phi} }.
\eea
Here $\eta(x) \in \R$ is part of the complex phase of $ \re^{\ii \eta(x)}\Phi$,
thus $D_\mu^{c,\Im}(\ii   \eta(x)) = \ii (D_\mu^c  \eta)$.
Now based on the same trick in \Eq{eq:D-phi}, we can absorb the U(1) gauge transformation by introducing the 1-form $A$ gauge field.
The $A$ transforms to $A'$ under the U(1) part of \Eq{eq:vector-gaugeA-w-matter}:
\bea
%\hspace{-10mm}
 A_{j }  &\to& A_{j }'=A_{j } +  \frac{1}{g}D_j^c   \eta,\\
%\hspace{-10mm}
 D_\mu^{A,c,\Im}  \log\Phi =\frac{ D_\mu^{A,c,\Im}\Phi }{\Phi}  &\to&
	D_\mu^{A',c,\Im}   \log\Phi  +  D_\mu^{A',c,\Im}(\ii   \eta(x))
%	= \frac{ D_\mu^{A,c,\Im}\Phi +  D_\mu^{c,\Im}(\ii   \eta(x)) \Phi}{\Phi} 
	=\frac{ D_\mu^{A',c,\Im}\Phi + \ii (D_\mu^c  \eta) \Phi}{\Phi} \nn\\
&&	=
	\frac{ 	(\prt_{\mu} - \ii g (A_{j } +  \frac{1}{g}D_j^c   \eta))\Phi -\ii g_c C_\mu \Phi_{\Im} + \ii (D_\mu^c  \eta) \Phi}{\Phi} 
=\frac{ D_\mu^{A,c,\Im}\Phi }{\Phi}. 
\quad\quad
\eea	
Here $D_\mu^{A',c,\Im}(\ii   \eta(x))=\ii (D_\mu^c  \eta)$ because $\eta$ is not charged under U(1) symmetry
but only the U(1) gauge parameter of $A$ itself. 
The above shows that
 $D_\mu^{A,c,\Im}  \log\Phi =\frac{ D_\mu^{A,c,\Im}\Phi }{\Phi}$ is a gauge invariant quantity under U(1) gauge transformation. 
 We can show that it is also gauge invariant quantity under the full O(2) gauge transformation (including  \Eq{eq:vector-gaugeA-w-matter} and the definition of
 $(\pm)^{\rtimes} \in \{+1,-1\}$ around \Eq{eq:vector-gaugeA-w-matter}):
 \bea
 A_{j }  &\to& A_{j }'=\re^{\ii \gamma_c(x)} A_{j } + (\pm)^{\rtimes} \frac{1}{g}(D_j^c ) ( \eta(x))  \eta,\\
 C_j &\to&  C_{j }'=C_j +\frac{1}{g_c} \prt_j \gamma_c(x), \\
 \Phi  &\to& \re^{ (\pm)^{\rtimes} \ii \eta(x)} \Phi_c =
  \re^{(\pm)^{\rtimes} \ii \eta(x)} (  \Phi_{\Re} + \ii \re^{\ii \gamma_c(x)} \Phi_{\Im}  )  =
 \left\{\begin{array}{l} 
 \re^{\ii \eta(x)} \Phi , \text{ if } \gamma_c \in  \pi \Z_{\text{even}},\\
 \re^{(\pm)^{\rtimes}\ii \eta(x)} \Phi^\dagger, \text{ if } \gamma_c \in  \pi \Z_{\text{odd}}.  \end{array}\right.\quad\quad
 \eea
Let us focus on the case $\Phi  \to \re^{ \ii \eta(x)} \Phi_c$ first, we have:
 \bea
 D_\mu^{A,c,\Im}  \log\Phi =\frac{ D_\mu^{A,c,\Im}\Phi }{\Phi}  &\to&
	D_\mu^{A',c',\Im}   \log\Phi_c  +  D_\mu^{A',c',\Im}(\ii   \eta(x))
	=\frac{ D_\mu^{A',c',\Im}\Phi_c + \ii (D_\mu^{c '} \eta) \Phi_c}{\Phi_c} \nn\\
	=&&\hspace{-10mm}
	\frac{ 	(\prt_{\mu} - \ii g (A_{j } +  \frac{1}{g}D_j^{c'}   \eta))\Phi_c -\ii g_c C'_\mu \re^{\ii \gamma_c} \Phi_{\Im} + \ii (D_\mu^{c'}  \eta) \Phi_c}{\Phi_c}
	\nn\\
	=&&\hspace{-10mm}
	\frac{ 	(\prt_{\mu} - \ii g (A_{j } +  \frac{1}{g}D_j^{c'}   \eta))\Phi_c -\ii g_c (C_j +\frac{1}{g_c} \prt_j \gamma_c) \re^{\ii \gamma_c}\Phi_{\Im} + \ii (D_\mu^{c'}  \eta) \Phi_c}{\Phi_c} 
	\nn\\
=&&\hspace{-10mm}
 \left\{\begin{array}{l} 
\frac{ D_\mu^{A,c,\Im}\Phi }{\Phi} , \text{ if } \gamma_c \in  \pi \Z_{\text{even}},\\
\frac{ D_\mu^{A,c,\Im}(\Phi^\dagger) }{\Phi^\dagger}, \text{ if } \gamma_c \in  \pi \Z_{\text{odd}}.  \end{array}\right.
\label{eq:DACphi-var-1}
\quad\quad
\eea	
Here some of the equalities hold when we focus on the leading order contribution for the gauge transformations.
Note that $\prt_{\mu}\Phi_c$ can contribute a $\prt_{\mu}(\re^{\ii \gamma_c} \Phi_{\Im})= {\ii (\prt_{\mu} \gamma_c)}\re^{\ii \gamma_c}\Phi_{\Im} + \dots$
that cancels with 
$-\ii g_c (\frac{1}{g_c} \prt_j \gamma_c) \re^{\ii \gamma_c}\Phi_{\Im}$.
We can define a complex conjugation operator of $D_\mu^{ A,c,\Im} $ as :
\bea
 D_\mu^{\dagger A,c,\Im} :=  (D_\mu^{ A,c,\Im} )^\dagger.
 \eea
 Similarly, we find under the  gauge transformation $\Phi  \to \re^{ \ii \eta(x)} \Phi_c$:
  \bea
 D_\mu^{\dagger A,c,\Im}  \log\Phi^\dagger
 &\to&\label{eq:DACphi-var-2}
 \left\{\begin{array}{l} 
\frac{ D_\mu^{\dagger A,c,\Im}\Phi^\dagger }{\Phi^\dagger} , \text{ if } \gamma_c \in  \pi \Z_{\text{even}},\\
\frac{ D_\mu^{\dagger A,c,\Im}(\Phi) }{\Phi}, \text{ if } \gamma_c \in  \pi \Z_{\text{odd}}.  \end{array}\right.
\\
 D_\mu^{A,c,\Im}  \log\Phi^\dagger
 &\to&\label{eq:DACphi-var-3}
 \left\{\begin{array}{l} 
\frac{ D_\mu^{A,c,\Im}\Phi^\dagger }{\Phi^\dagger} , \text{ if } \gamma_c \in  \pi \Z_{\text{even}},\\
\frac{ D_\mu^{A,c,\Im}(\Phi) }{\Phi}, \text{ if } \gamma_c \in  \pi \Z_{\text{odd}}.  \end{array}\right.\\
 D_\mu^{\dagger A,c,\Im}  \log\Phi
 &\to&\label{eq:DACphi-var-4}
 \left\{\begin{array}{l} 
\frac{ D_\mu^{\dagger A,c,\Im}\Phi }{\Phi} , \text{ if } \gamma_c \in  \pi \Z_{\text{even}},\\
\frac{ D_\mu^{\dagger A,c,\Im}(\Phi^\dagger) }{\Phi^\dagger}, \text{ if } \gamma_c \in  \pi \Z_{\text{odd}}.  \end{array}\right.
 \eea
  Similar discussions follow after taking the factor $ (\pm)^{\rtimes}$ into the account.
 Since the denominators in
 \Eq{eq:DACphi-var-1}, \Eq{eq:DACphi-var-2},
  \Eq{eq:DACphi-var-3}, and \Eq{eq:DACphi-var-4}
are all \emph{gauge-covariant} or \emph{complex conjugation gauge-covariant}:
\bea
 \Phi  \to \re^{ (\pm)^{\rtimes} \ii \eta(x)} \Phi_c, 
 \eea
 this means that the numerators are
 also \emph{gauge-covariant} or \emph{complex conjugation gauge-covariant}.
 We can construct the gauge invariant quantity via pairing the 
 \emph{gauge-covariant} term with its complex conjugation,
 and pairing 
 the \emph{complex conjugation gauge-covariant} term also with its complex conjugation.
 So we obtain:\footnote{Here we are allowed to flip the sign $C \to -C$ since $C$ is only a $\Z_2$ gauge field.
 More precisely, $\oint C = \frac{2\pi}{2} \Z = \pi \Z \mod 2 \pi$,
 while $\oint C = - \oint C \mod 2 \pi$.
 It may also look peculiar that the particle $\Phi$ and anti-particle 
 $\Phi^\dagger$ both couples to the gauge field $A$ with both $\pm 1$ couplings.
 However, we may comfort the readers by reminding the fact that the particle-hole conjugation symmetry $\Z_2^C$ is
 already gauged, thus $\Z_2^C$ gauge field couples to both the particle $\Phi$ and anti-particle 
 $\Phi^\dagger$, via the $\Phi_{\Im}$ part. 
Furthermore, the $\Z_2^C$ gauge field can flip the sign of their U(1) gauge charge $+ 1 \leftrightarrow -1$.
This seems to suggest that
particle and anti-particle may share part of the degree of freedom.
This reminds us the famous fact 
that Majorana fermion has the particle and anti-particle identified as the same, although we should beware that our particle $\Phi$ is bosonic instead.
 \label{ft:flip-C}}
 \bea
&&
\label{eq:gauged-matter-degree-0-O(2)-L}
(D_{\mu}^{A,c,\Im} \Phi) ({D^{\dagger \mu}_{A,c,\Im}} \Phi^\dagger)
+(D_{\mu}^{\dagger A,c,\Im} \Phi) ({D^{\mu}_{A,c,\Im}} \Phi^\dagger)\\
&&=
\big((\prt_{\mu} - \ii g A_{\mu})\Phi -\ii g_c C_\mu \Phi_{\Im}
\big)
\big((\prt^{\mu} + \ii g A^{\mu})\Phi^\dagger +\ii g_c C^\mu \Phi_{\Im}
\big)\nn
\\
&&+
\big((\prt_{\mu} + \ii g A_{\mu})\Phi +\ii g_c C_\mu \Phi_{\Im}
\big)
\big((\prt^{\mu} - \ii g A^{\mu})\Phi^\dagger -\ii g_c C^\mu \Phi_{\Im}
\big).\nn
\eea

\paragraph{Approach 2: Yang-Mills method}

Let us cross-check the above result from a more conventional Yang-Mills method \cite{PhysRev96191YM1954}.
To start with, we observe that
$(\dd + X) \begin{pmatrix}
\Phi_\Re\\
\Phi_{\Im}
\end{pmatrix} $ must be gauge covariant under the gauge transformation. Because
under a generic gauge transformation $V_{\tO(2)}$, 
it demands
$(\dd + X) \to (V_{\tO(2)} (\dd + X) V_{\tO(2)}^{-1})$
and 
$\begin{pmatrix}
\Phi_\Re\\
\Phi_{\Im}
\end{pmatrix}
\to 
(V_{\tO(2)}   \begin{pmatrix}
\Phi_\Re\\
\Phi_{\Im}
\end{pmatrix})$,
we show the gauge covariance of
\bea
(\dd + X) \begin{pmatrix}
\Phi_\Re\\
\Phi_{\Im}
\end{pmatrix}
\to
(V_{\tO(2)} (\dd + X) V_{\tO(2)}^{-1}) (V_{\tO(2)}   \begin{pmatrix}
\Phi_\Re\\
\Phi_{\Im}
\end{pmatrix})=
V_{\tO(2)}
(\dd + X) \begin{pmatrix}
\Phi_\Re\\
\Phi_{\Im}
\end{pmatrix}.
\eea
Since all components of $A,C,\Phi_{\Re}, \Phi_{\Im} \in \R$ are reals, we can pair the gauge covariant term with its transpose (the Hodge dual $\star$) to obtain
a gauge invariant Lagrangian term (again see footnote \ref{ft:flip-C})
\bea
&&(\dd + X) \begin{pmatrix}
\Phi_\Re\\
\Phi_{\Im}
\end{pmatrix} \wedge \star (\dd + X) \begin{pmatrix}
\Phi_\Re\\
\Phi_{\Im}
\end{pmatrix}\\
&&=
 |\begin{pmatrix}
\prt_\mu &  g A_\mu\\
-  g A_\mu & \prt_\mu - g_c  C_\mu
\end{pmatrix}\begin{pmatrix}
\Phi_\Re\\
\Phi_{\Im}
\end{pmatrix}|^2
=
|\begin{pmatrix}
\prt_\mu \Phi_\Re + g A_\mu\Phi_{\Im}\\
\prt_\mu \Phi_{\Im} - g A_\mu\Phi_\Re
   -  g_c C_\mu
\Phi_{\Im}
\end{pmatrix}|^2 \nn\\
&&
=(\prt_\mu \Phi_\Re + g A_\mu\Phi_{\Im})^2
+(\prt_\mu \Phi_{\Im} - g A_\mu\Phi_\Re
 - g_c  C_\mu
\Phi_{\Im})^2 \nn\\
&&=
(\prt_\mu \Phi_\Re + g A_\mu\Phi_{\Im})^2
+(\prt_\mu \Phi_{\Im} - g A_\mu\Phi_\Re)^2
 + ( g_c C_\mu
\Phi_{\Im})^2
-2(\prt_\mu \Phi_{\Im} - g A_\mu\Phi_\Re)
(g_c C^\mu
\Phi_{\Im}) \nn\\
&&
=D_{\mu}^A \Phi D^{\dagger A,\mu} \Phi^\dagger
 + ( g_c C_\mu
\Phi_{\Im})^2
-2(\prt_\mu \Phi_{\Im} - g A_\mu\Phi_\Re)
( g_c C^\mu
\Phi_{\Im}) %\nn\\
%&&
=
(D_{\mu}^{A,c,\Im} \Phi) ({D^{\dagger \mu}_{A,c,\Im}} \Phi^\dagger).
\eea
Thus we show (again see footnote \ref{ft:flip-C})
\bea
\boxed{
(D_{\mu}^{A,c,\Im} \Phi) ({D^{\dagger \mu}_{A,c,\Im}} \Phi^\dagger)=
(\dd + X) \begin{pmatrix}
\Phi_\Re\\
\Phi_{\Im}
\end{pmatrix} \wedge \star (\dd + X) \begin{pmatrix}
\Phi_\Re\\
\Phi_{\Im}
\end{pmatrix}}
.
\eea
Similarly (again see footnote \ref{ft:flip-C}),
\bea
\boxed{
(D_{\mu}^{A,c,\Im} \Phi^\dagger) ({D^{\dagger \mu}_{A,c,\Im}} \Phi )=
(\dd - X) \begin{pmatrix}
\Phi_\Re\\
\Phi_{\Im}
\end{pmatrix} \wedge \star (\dd - X) \begin{pmatrix}
\Phi_\Re\\
\Phi_{\Im}
\end{pmatrix}
=
(\dd + X) \begin{pmatrix}
\Phi_\Re\\
-\Phi_{\Im}
\end{pmatrix} \wedge \star (\dd + X) \begin{pmatrix}
\Phi_\Re\\
-\Phi_{\Im}
\end{pmatrix}}
.\quad
\eea
Thus we can construct an
O(2) gauged matter field theory contains a Lagrangian term
\Eq{eq:gauged-matter-degree-0-O(2)-L} and a potential $ V(|\Phi|^2)$ as:
\bea \label{eq:gauged-matter-degree-0-O(2)-L-V}
\boxed{
(D_{\mu}^{A,c,\Im} \Phi) ({D^{\dagger \mu}_{ A,c,\Im}} \Phi^\dagger)
+(D_{\mu}^{\dagger A,c,\Im} \Phi) ({D^{\mu}_{A,c,\Im}} \Phi^\dagger)+ V(|\Phi|^2)}.
\eea
The theory contains particle $\Phi$ and anti-particle $\Phi^\dagger$ pair together 
in an intricate way because the  
particle-hole $\Z_2^C$ symmetry $\Phi \overset{\Z_2^C}{\longleftrightarrow} \Phi^\dagger$ is also dynamically gauged.
If the particle $\Phi$ has a gauge charge-1, then the
anti-particle $\Phi^\dagger$ has a gauge charge-$(-1)$ under the U(1) gauge group. (However, see also footnote \ref{ft:flip-C})

Follow \cite{Wang2019aiq1909.13879, WXY2-1910-1911.01804}, we can consider the $N$-layers generalization of the theories
with $(\Z_2^C)^N$ gauged,
also by including the O(2)-Yang Mills kinetic term \Eq{eq:rank-2-nAb-F-L} and the level-2 BF theory
into the O(2) gauge matter theory \Eq{eq:gauged-matter-degree-0-O(2)-L-V},
we are allowed to introduce the twisted cocycle 
$\omega_{d+1} \in \rH^{d+1}((\Z_2^C)^N, \R/\Z)$ from a group cohomology data \cite{DijkgraafWitten1989pz1990} to specify
the interlayer interactions between $N$-layers. We can write down a schematic path integral:
\bea
&&\bZ_{\text{rk-2-NAb-$\Phi$}} =\int 
(\prod_{I=1}^{N} [\cD A_I][\cD B_I]  [\cD C_I][\cD \Phi_I][\cD \Phi_I^\dagger])\exp(\ii \int_{M^{d+1}} \dd^{d+1}x \Big(  \sum_{I=1}^{N}\big(  |\hat F^{c,I}_{\mu\nu}|^2 \nn\\
&&+(D_{\mu}^{A,c,\Im} \Phi_I ) ({D^{\dagger \mu}_{A,c,\Im}} \Phi_I^\dagger)
+(D_{\mu}^{\dagger A,c,\Im} \Phi_I) ({D^{\mu}_{A,c,\Im}} \Phi_I^\dagger) +V(\{ |\Phi_I|^2\}) \big) + \frac{ 2}{2 \pi} B_I \dd C_I\Big))
\cdot \omega_{d+1}(\{C_I\}).\quad\quad\quad
\eea

\subsection{Degree-1 polynomial symmetry to Pretko's field theory and non-abelian generalization}

\label{sec:Degree-1-Pretko's}

\subsubsection{Global-covariant 2-derivative}

Now we construct a field theory that preserves a {degree-1 polynomial} symmetry with a polynomial $Q(x)=(\Lambda_{k}  x_k + \Lambda_0)$.
A {degree-1 polynomial} symmetry transforms $\Phi$ and $\log\Phi$ as
\bea
\Phi &\to&\re^{\ii Q(x)}\Phi =   \re^{\ii (\Lambda_{k}  x_k + \Lambda_0)}\Phi,\\
\log \Phi &\to& \log\Phi+ \ii Q(x) = \log\Phi+ {\ii (\Lambda_{i}  x_i + \Lambda_0)}.
\eea
Take $\prt_{x_i}\prt_{x_j}:= \prt_{i}\prt_{j}$ on both sides, we construct a globally invariant term, 
\bea
\prt_{i}\prt_{j}\log\Phi 
		 \to \prt_{i}\prt_{j}\log\Phi.
\eea
We also define 
\bea \label{eq:ddlogphi}
\prt_{i}\prt_{j}\log\Phi :=
		\frac{P_{i,j}(\Phi,\prt\Phi,\prt^2\Phi)}{\Phi^2}= \frac{{\Phi \prt_{i} \prt_{j} \Phi -( \prt_{i} \Phi)( \prt_{j} \Phi) }}{\Phi^2}.
\eea
Under $\Phi\to  \re^{\ii Q(x)}\Phi$, since the denominator $\Phi^2 \to \re^{\ii 2Q(x)}\Phi^2$, so does
the numerator ${P_{i,j}(\Phi,\prt\Phi,\prt^2\Phi)} \to \re^{\ii 2Q(x)} {P_{i,j}(\Phi,\prt\Phi,\prt^2\Phi)} $ which we name 
\bea
{P_{i,j}(\Phi,\prt\Phi,\prt^2\Phi)}:= {{\Phi \prt_{i} \prt_{j} \Phi -( \prt_{i} \Phi)( \prt_{j} \Phi) }}
\eea
 as a global-covariant 2-derivative term, 
in order to maintain the $\prt_{i}\prt_{j}\log\Phi$ to be invariant.
The gauge-invariant Lagrangian contains 
\bea
|{P_{i,j}}|^2 + V(|\Phi|^2) := {P_{i,j}}(\Phi){P^{i,j}}(\Phi^\dagger) + V(|\Phi|^2)
= ({{\Phi \prt_{i} \prt_{j} \Phi - \prt_{i} \Phi \prt_{j} \Phi }} )
 ({{\Phi^\dagger \prt^{i} \prt^{j} \Phi^\dagger - \prt^{i} \Phi \prt^{j} \Phi^\dagger }} )
 + V(|\Phi|^2). \quad
\eea

%\cblue{JW-Xu-Yau (1911.01804)} reproduces Pretko (2018).
In this way, based on the systematic method of  \Refe{WXY2-1910-1911.01804}, we can re-derive a Lagrangian formulation of
Pretko in 2018 \cite{Pretko2018jbi1807.11479}, which are recently revisited in \cite{Gromov2018nbv1812.05104, SeibergF2019} and \cite{Wang2019aiq1909.13879} from other field theory perspectives. 

\subsubsection{Gauge-covariant 2-derivative}
\label{sec:Gauge-covariant-2-derivative}

To gauge a {degree-0 polynomial} symmetry,	we rewrite $Q(x)$ as a local gauge parameter $\eta(x)$,
\bea
\Phi &\to&  \re^{\ii \eta(x)}\Phi ,\\
\prt_{i}\prt_{j}\log\Phi &\to& \prt_{i}\prt_{j}\log\Phi  +  \ii \prt_{i}\prt_{j}\eta(x).
\eea
Then $\prt_{i} \prt_{j}\log\Phi $ is no longer an invariant term.
This implies that we can write a new gauge-covariant operator $D_{i,j}[\{\Phi \}]$ via
 via
combining $P_{i,j}$ and $A_{i,j}$:
	\bea
	P_{i,j}(\Phi,\prt\Phi,\prt^2\Phi) &:=& ({{\Phi \prt_{i} \prt_{j} \Phi -( \prt_{i} \Phi)( \prt_{j} \Phi) }}) \to\re^{\ii 2 \eta(x)} (P_{i,j}(\Phi,\prt\Phi,\prt^2\Phi)
	+\ii \prt_{i} \prt_{j}\eta(x)). \label{eq:Pij-gauge-var}\\
        	A_{i,j}  &\to &A_{i,j} +\frac{1}{g} \prt_{i} \prt_{j} \eta.  \label{eq:Aij-gauge-var}\\
	        D_{i,j}^A[\{\Phi \}] &:=&P_{i ,j}(\Phi,\prt\Phi,\prt^2\Phi) - \ii g A_{i,j} \Phi^2 = ({{\Phi \prt_{i} \prt_{j} \Phi -( \prt_{i} \Phi)( \prt_{j} \Phi) }} - \ii g A_{i,j} \Phi^2).\\
	        D_{i,j}^A[\{\Phi \}] &\to & \re^{\ii  2\eta(x)} D_{i,j}^A[\{\Phi \}]. 
	\eea
	We shall call $D_{i,j}^A[\{\Phi \}]$ a gauge-covariant 2-derivative term.\footnote{
	Since $D_{i,j}^A[\{\Phi \}] \to  \re^{\ii  2\eta(x)} D_{i,j}^A[\{\Phi \}]$ with a covariant factor of power 2 as 
	$\re^{\ii  2\eta(x)}$, we may call this as ``2-covariant'' for the convenience. \label{footnote:2-covariant}
	}
 So a gauge invariant Lagrangian term can be obtained by complex-conjugate pairing the gauge-covariant operator as
\bea
\left| D_{i,j}^A[\{\Phi \}] \right|^2  + V(|\Phi|^2)= D_{i,j}^A[\{\Phi \}]  D^{\dagger i,j}_A[\{\Phi^\dagger \}] 
 + V(|\Phi|^2).
\eea
Thus we also reproduce Pretko's abelian gauge theory \cite{Pretko2018jbi1807.11479}.

\subsubsection{Gauge-covariant non-abelian $[{ \U(1)_{x_{(d)} }  \rtimes \Z_2^C}]$ rank-3 field strength}

Follow  \Refe{Wang2019aiq1909.13879, WXY2-1910-1911.01804}, 
we define a non-abelian rank-3 field strength
\bea
{\hat F^c_{\mu \nu \xi} :=D_\mu^c A_{\nu \xi} -D_{\nu }^c A_{\mu \xi }
:=(\prt_\mu - \ii g_c C_\mu) A_{\nu \xi} -(\prt_{\nu} - \ii g_c C_\nu ) A_{\mu \xi}}. 
\eea
Under the gauge transformation
 \bea \label{eq:vector-gauge-charge}
   A_{\mu \nu}  &\to& e^{\ii \gamma_c(x)} A_{\mu \nu}  +    (\pm)^{\rtimes} \frac{1}{2g}(D_\mu^c D_\nu^c+ D_\nu^c D_\mu^c) ( \eta_v(x)) \nn\\
     & =&\left\{\begin{array}{l} 
   V_{\Z_2} V_{\U(1)}A_{\mu \nu}= \re^{\ii \gamma_c(x)} A_{\mu \nu} +  \frac{1}{2g}(D_\mu^c D_\nu^c+ D_\nu^c D_\mu^c) \eta_v, \\
   V_{\U(1)}V_{\Z_2}A_{\mu \nu}= \re^{\ii \gamma_c(x)} (A_{\mu \nu} +  \frac{1}{2g}(D_\mu^c D_\nu^c+ D_\nu^c D_\mu^c)\eta_v) \end{array}\right..
   \nn\\
   C_\nu &\to& C_\nu +\frac{1}{g_c} \prt_\nu \gamma_c(x). \nn
   \eea
   we can again show ${\hat F^c_{\mu \nu \xi}}$ is gauge-covariant:
   \bea
   {\hat F^c_{\mu \nu \xi}} &\to&
    e^{\ii \gamma_c(x)} \hat F^c_{\mu \nu \xi}
 +  \frac{1}{2g}\Big( D_\mu^c(D_\nu^c D_\xi^c+ D_\xi^c D_\nu^c)  
 -
 D_\nu^c
 (D_\mu^c D_\xi^c+ D_\xi^c D_\mu^c)
 \Big)\eta_v=   e^{\ii \gamma_c(x)} \hat F^c_{\mu \nu \xi}.
\eea
%%%%%%%%%
This is true because under the locally flat $\dd C=0$ condition, we had derived 
$(D_\mu^c D_\nu^c- D_\nu^c D_\mu^c)=0$
in \Eq{eq:DD-DD}, furthermore
\bea
   &&\hspace{-10mm}
   {\Big( D_\mu^c(D_\nu^c D_\xi^c+ D_\xi^c D_\nu^c)  
 -
 D_\nu^c
 (D_\mu^c D_\xi^c+ D_\xi^c D_\mu^c)
 \Big)}
 =
 (D_\mu^c D_\nu^c- D_\nu^c D_\mu^c)
 D_\xi^c
 +(D_\mu^c
 D_\xi^c D_\nu^c
 - D_\nu^c D_\xi^c D_\mu^c)\nn\\
&& \hspace{-10mm} =
 (D_\mu^c
 D_\xi^c D_\nu^c
 - D_\nu^c D_\xi^c D_\mu^c)
  =
 (\prt_\mu - \ii g_c C_\mu)(\prt_{\xi} - \ii g_c C_{\xi} )(\prt_{\nu} - \ii g_c C_\nu )
 - (\prt_{\nu} - \ii g_c C_\nu )(\prt_{\xi} - \ii g_c C_{\xi} ) (\prt_\mu - \ii g_c C_\mu) \nn\\
 &&  =
 (\prt_\mu - \ii g_c C_\mu) 
 (\prt_{\xi}\prt_{\nu} - \ii g_c C_{\xi}\prt_{\nu}  - \ii g_c   (\prt_{\xi}C_\nu) - \ii g_c C_\nu  \prt_{\xi} -  g_c^2C_{\xi} C_\nu ) 
 - (\mu \leftrightarrow \nu)\nn\\
 &&  = - \ii g_c   (\prt_{\xi}C_\nu){(\prt_\mu- \ii g_c C_\mu) }
   - \ii g_c  {(\prt_\mu C_{\xi}){(\prt_\nu- \ii g_c C_\nu) } }
   - (\mu \leftrightarrow \nu) \nn\\
    &&  =  \ii g_c \big( (\dd C)_{\nu \xi}   {(\prt_\mu- \ii g_c C_\mu) } - (\dd C)_{ \xi \mu} {(\prt_\nu- \ii g_c C_\nu) } \big) \vert_{\dd C=0}=0
\eea
The $(\mu \leftrightarrow \nu)$ are the term exchanging $\mu$ and $\nu$ respect to the previous term.
The non-abelian field strength has firstly appeared in  \Refe{Wang2019aiq1909.13879, WXY2-1910-1911.01804}.
The gauge-invariant non-abelian gauge field kinetic Lagrangian term corresponds to:
 \bea
 \label{eq:rank-3-nAb-F-L}
 | {\hat F^c_{\mu \nu \xi}}|^2:= {\hat F^c_{\mu \nu \xi}} {\hat F^{\dagger c \mu \nu \xi}}
 \eea
\subsubsection{Non-abelian $[{ \U(1)_{x_{(d)} }  \rtimes \Z_2^C}]$ gauged matter: Polynomial invariant method}
\label{sec:degree-1-nab-gauged-M}

Follow the first approach in \Sec{sec:degree-0-nab-gauged-M}, we construct non-abelian gauged matter theory.
By generalizing \Eq{eq:ddlogphi},
we consider
\bea
&& \hspace{-50pt}
D_\mu^{c,\Im}  D_\nu^{c,\Im}\log\Phi =
D_\mu^{c,\Im}  ( (D_\nu^{c,\Im} \Phi) \Phi^{-1} )=D_\mu^{c,\Im}  ( (D_\nu^{c,\Im} \Phi) \frac{\Phi^\dagger}{|\Phi|^2} )\nn\\
&& \hspace{-50pt}
= (  \frac{\Phi (D_\mu^{c,\Im} D_\nu^{c,\Im} \Phi)}{\Phi^2} ) + (D_\nu^{c,\Im} \Phi) (D_\mu^{c,\Im} \Phi^{-1}) \nn\\
&& \hspace{-50pt}
= (  \frac{\Phi (D_\mu^{c,\Im} D_\nu^{c,\Im} \Phi)}{\Phi^2} ) + (D_\nu^{c,\Im} \Phi) (D_\mu^{c,\Im} \frac{\Phi_{\Re} - \ii \Phi_{\Im} }{\Phi_{\Re}^2+ \Phi_{\Im} ^2}) \nn\\
%&& \hspace{-50pt}
%= (  \frac{\Phi (D_\mu^{c,\Im} D_\nu^{c,\Im} \Phi)}{\Phi^2} ) +  
%\frac{-(D_\mu^{c,\Im}  \Phi)(D_\nu^{c,\Im} \Phi) + g_c C_\mu (\frac{ - 2 \ii \Phi_{\Im}^2   }{\Phi^\dagger} ) (D_\nu^{c,\Im} \Phi)}{\Phi^2} 
%\\
&& \hspace{-50pt}
= (  \frac{\Phi (D_\mu^{c,\Im} D_\nu^{c,\Im} \Phi)}{\Phi^2} ) +  
\frac{-(D_\mu^{c,\Im}  \Phi)(D_\nu^{c,\Im} \Phi) + g_c C_\mu (\frac{ - 2 \ii \Phi_{\Im}^2  \Phi  }{|\Phi|^2} ) (D_\nu^{c,\Im} \Phi)}{\Phi^2}. 
\label{eq:DcDclogphi}
\eea
In \Sec{sec:Gauge-covariant-2-derivative}, we had learned that for the abelian gauge sector, we require to introduce a symmetric tensor gauge field $A_{\mu,\nu}$
in order to cancel the gauge transformation $\prt_{i} \prt_{j} \eta$ between \Eq{eq:Pij-gauge-var} and  \Eq{eq:Aij-gauge-var}.
Thus, we also symmetrize the above equation\footnote{Below we use the notation $\{ J_1,J_2\}_+ := J_1 J_2+ J_2J_1$ to define the anti-commutator, e.g.
\bea
\{D_\mu^{c,\Im} , D_\nu^{c,\Im} \}_+:= { {D_\mu^{c,\Im} D_\nu^{c,\Im}}+ {D_\nu^{c,\Im} D_\mu^{c,\Im}} }.
\eea
} 
 in order to naturally couple to a symmetric tensor gauge field later:
\bea
&& \hspace{-50pt}
\frac{\{D_\mu^{c,\Im} , D_\nu^{c,\Im} \}_+}{2} \log\Phi
=
\frac{(\Phi \frac{\{D_\mu^{c,\Im} , D_\nu^{c,\Im} \}_+}{2} \Phi - D_\mu^{c,\Im} \Phi D_\nu^{c,\Im} \Phi 
+ g_c  (\frac{ -  \ii \Phi_{\Im}^2   }{\Phi^\dagger} )  ((C_\mu D_\nu^{c,\Im} +C_\nu D_\mu^{c,\Im} )  \Phi)  
)}{\Phi^2}. 
\label{eq:DmuDnucImlogPhi}
\eea
The $[{ \U(1)_{x_{(d)} }  \rtimes \Z_2^C}]$ gauged transformations are:
 \bea \label{eq:degree1-A-gauge-trans}
   A_{\mu \nu}  &\to&   A_{\mu \nu}'= e^{\ii \gamma_c(x)} A_{\mu \nu}  +    (\pm)^{\rtimes} \frac{1}{2g}(D_\mu^c D_\nu^c+ D_\nu^c D_\mu^c) ( \eta_v(x))
   = e^{\ii \gamma_c(x)} A_{\mu \nu}  +    (\pm)^{\rtimes} \frac{1}{2g}(\{ D_\mu^c, D_\nu^c\}_+  \eta_v(x)). \quad\quad\quad\\
   \label{eq:degree1-C-gauge-trans}
 C_j &\to&  C_{j }'=C_j +\frac{1}{g_c} \prt_j \gamma_c(x). \\
 \label{eq:degree1-Phi-gauge-trans}
 \Phi  &\to& \re^{ (\pm)^{\rtimes} \ii \eta(x)} \Phi_c =
  \re^{(\pm)^{\rtimes} \ii \eta(x)} (  \Phi_{\Re} + \ii \re^{\ii \gamma_c(x)} \Phi_{\Im}  )  =
 \left\{\begin{array}{l} 
 \re^{\ii \eta(x)} \Phi , \text{ if } \gamma_c \in  \pi \Z_{\text{even}},\\
 \re^{(\pm)^{\rtimes}\ii \eta(x)} \Phi^\dagger, \text{ if } \gamma_c \in  \pi \Z_{\text{odd}}.  \end{array}\right.\quad\quad
 \eea
 Under the $[{ \U(1)_{x_{(d)} }  \rtimes \Z_2^C}]$ gauged transformation,
$\frac{\{D_\mu^{c,\Im} , D_\nu^{c,\Im} \}_+}{2} \log\Phi $ is not gauge invariant,
thus the numerator in \Eq{eq:DmuDnucImlogPhi}, 
${(\Phi \frac{\{D_\mu^{c,\Im} , D_\nu^{c,\Im} \}_+}{2} \Phi - D_\mu^{c,\Im} \Phi D_\nu^{c,\Im} \Phi 
+ g_c  (\frac{ -  \ii \Phi_{\Im}^2   }{\Phi^\dagger} )  ((C_\mu D_\nu^{c,\Im} +C_\nu D_\mu^{c,\Im} )  \Phi)  
)}$,
is also not gauge-covariant.
This result is what we should expect,
because there is a variant term $\ii \frac{\{D_\mu^{c,\Im} , D_\nu^{c,\Im} \}_+}{2}\eta$ 
in this non-abelian theory, similar to the variant term $\ii \prt_{i} \prt_{j}\eta(x)$ appears in the abelian version of \Eq{eq:Pij-gauge-var}.
However, the symmetric tensor gauge field can cancel such a gauge variant term exactly. 

So we define a new non-abelian gauge-covariant 2-derivative for this non-abelian theory
(generalizing the abelian case in \Sec{sec:Gauge-covariant-2-derivative}):
 \bea  \hspace{-10pt} \label{eq:Gauge-covariant-2-derivative-nab-1}
\boxed{D_{\mu,\nu}^{A,c,\Im}[\{\Phi \}] := {(\Phi \frac{\{D_\mu^{c,\Im} , D_\nu^{c,\Im} \}_+}{2} \Phi - D_\mu^{c,\Im} \Phi D_\nu^{c,\Im} \Phi 
+ g_c  (\frac{ -  \ii \Phi_{\Im}^2   }{\Phi^\dagger} )  ((C_\mu D_\nu^{c,\Im} +C_\nu D_\mu^{c,\Im} )  \Phi)  
- \ii g  A^{}_{\mu \nu} {\Phi^2}
)}}.\quad\quad
\eea
Similarly, we have:
 \bea  &&\hspace{-40pt} \label{eq:Gauge-covariant-2-derivative-nab-2}
D_{\mu,\nu}^{\dagger A,c,\Im}[\{\Phi^\dagger \}]:={(\Phi^\dagger \frac{\{D_\mu^{c,\Im \dagger} , D_\nu^{c,\Im \dagger} \}_+}{2} \Phi^\dagger - D_\mu^{c,\Im \dagger} \Phi^\dagger D_\nu^{c,\Im \dagger} \Phi^\dagger 
+ g_c  (\frac{ +  \ii \Phi_{\Im}^2   }{\Phi} )  ((C_\mu D_\nu^{c,\Im \dagger}  +C_\nu D_\mu^{c,\Im \dagger} )  \Phi^\dagger)  
+ \ii g A^{}_{\mu \nu} \Phi^{\dagger 2})},\nn\\
  &&\hspace{-40pt} \label{eq:Gauge-covariant-2-derivative-nab-3}
D_{\mu,\nu}^{\dagger A,c,\Im}[\{\Phi \}] := {{(\Phi \frac{\{D_\mu^{c,\Im \dagger} , D_\nu^{c,\Im \dagger} \}_+}{2} \Phi - D_\mu^{c,\Im \dagger} \Phi D_\nu^{c,\Im \dagger} \Phi 
+ g_c  (\frac{ +  \ii \Phi_{\Im}^2   }{\Phi^\dagger} )  ((C_\mu D_\nu^{c,\Im \dagger} +C_\nu D_\mu^{c,\Im \dagger} )  \Phi)  
- \ii g A^{{}}_{\mu \nu} \Phi^2)}},\nn\\
  &&\hspace{-40pt} \label{eq:Gauge-covariant-2-derivative-nab-4}
D_{\mu,\nu}^{A,c,\Im}[\{\Phi^\dagger \}] :={(\Phi^\dagger \frac{\{D_\mu^{c,\Im } , D_\nu^{c,\Im } \}_+}{2} \Phi^\dagger - D_\mu^{c,\Im } \Phi^\dagger D_\nu^{c,\Im } \Phi^\dagger 
+ g_c  (\frac{ -  \ii \Phi_{\Im}^2   }{\Phi} )  ((C_\mu D_\nu^{c,\Im }  +C_\nu D_\mu^{c,\Im } )  \Phi^\dagger)  
+ \ii g A^{}_{\mu \nu} \Phi^{\dagger 2})}.
\eea
Such that \Eq{eq:Gauge-covariant-2-derivative-nab-1} to \Eq{eq:Gauge-covariant-2-derivative-nab-4} are gauge 2-covariant  (see footnote \ref{footnote:2-covariant})
or gauge 2-covariant to its complex conjugate field $\Phi^\dagger$ under the 
gauge transformations \Eq{eq:degree1-A-gauge-trans}, \Eq{eq:degree1-C-gauge-trans} and \Eq{eq:degree1-Phi-gauge-trans}:
  \bea
D_{\mu,\nu}^{A,c,\Im}[\{\Phi \}] 
 &\to&\label{eq:DDACphi-var-1}
 \left\{\begin{array}{l} 
\re^{\ii  2\eta(x)} D_{\mu,\nu}^{A,c,\Im}[\{\Phi \}]  , \text{ if } \gamma_c \in  \pi \Z_{\text{even}},\\
\re^{\ii  (\pm)^{\rtimes} 2\eta(x)} D_{\mu,\nu}^{A,c,\Im}[\{\Phi^\dagger \}] , \text{ if } \gamma_c \in  \pi \Z_{\text{odd}}.  \end{array}\right.
\\
D_{\mu,\nu}^{\dagger A,c,\Im}[\{\Phi^\dagger \}] 
 &\to&\label{eq:DDACphi-var-2}
 \left\{\begin{array}{l} 
\re^{-\ii  2\eta(x)} D_{\mu,\nu}^{\dagger A,c,\Im}[\{\Phi^\dagger \}]  , \text{ if } \gamma_c \in  \pi \Z_{\text{even}},\\
\re^{-\ii  (\pm)^{\rtimes} 2\eta(x)} D_{\mu,\nu}^{\dagger A,c,\Im}[\{\Phi \}] , \text{ if } \gamma_c \in  \pi \Z_{\text{odd}}.  \end{array}\right.
\\
D_{\mu,\nu}^{\dagger A,c,\Im}[\{\Phi \}] 
 &\to&\label{eq:DDACphi-var-3}
 \left\{\begin{array}{l} 
\re^{\ii  2\eta(x)}  D_{\mu,\nu}^{\dagger A,c,\Im}[\{\Phi \}] , \text{ if } \gamma_c \in  \pi \Z_{\text{even}},\\
\re^{\ii  (\pm)^{\rtimes} 2\eta(x)} D_{\mu,\nu}^{\dagger A,c,\Im}[\{\Phi^\dagger \}] , \text{ if } \gamma_c \in  \pi \Z_{\text{odd}}.  \end{array}\right.
\\
D_{\mu,\nu}^{A,c,\Im}[\{\Phi^\dagger \}] 
 &\to&\label{eq:DDACphi-var-4}
 \left\{\begin{array}{l} 
\re^{-\ii  2\eta(x)} D_{\mu,\nu}^{A,c,\Im}[\{\Phi^\dagger \}] , \text{ if } \gamma_c \in  \pi \Z_{\text{even}},\\
\re^{-\ii  (\pm)^{\rtimes} 2\eta(x)} D_{\mu,\nu}^{A,c,\Im}[\{\Phi \}] , \text{ if } \gamma_c \in  \pi \Z_{\text{odd}}.  \end{array}\right.
 \eea

Follow \cite{Wang2019aiq1909.13879, WXY2-1910-1911.01804}, we can consider the $N$-layers generalization of the theories
with $(\Z_2^C)^N$ gauged,
also by including the $[{ \U(1)_{x_{(d)} }  \rtimes \Z_2^C}]$-gauge kinetic term \Eq{eq:rank-3-nAb-F-L} and the level-2 BF theory
into the gauge matter theory,
again we are allowed to introduce the twisted cocycle 
$\omega_{d+1} \in \rH^{d+1}((\Z_2^C)^N, \R/\Z)$ from a group cohomology data \cite{DijkgraafWitten1989pz1990} to specify
the interlayer interactions between $N$-layers. We can write down a schematic path integral (see also footnote \ref{ft:flip-C}):
\bea
&&\bZ_{\text{rk-3-NAb-$\Phi$}} =\int 
(\prod_{I=1}^{N} [\cD A_I][\cD B_I]  [\cD C_I][\cD \Phi_I][\cD \Phi_I^\dagger])\exp(\ii \int_{M^{d+1}}  \sum_{I=1}^{N} \Big(  \dd^{d+1}x\big(  |\hat F^{c,I}_{\mu\nu\xi}|^2 \\
&&+D_{\mu,\nu}^{A,c,\Im}[\{\Phi_I \}]  D^{\dagger {\mu,\nu}}_{ A,c,\Im}[\{\Phi_I^\dagger \}] 
+D_{\mu,\nu}^{\dagger A,c,\Im}[\{\Phi_I \}]  D^{\mu,\nu}_{A,c,\Im}[\{\Phi_I^\dagger \}]  +V(\{ |\Phi_I|^2\})\big)  + \frac{ 2}{2 \pi} B_I \dd C_I\Big) )
\cdot \omega_{d+1}(\{C_I\}).\nn
\eea
The new ingredient in our present work beyond the previous \Refe{Wang2019aiq1909.13879, WXY2-1910-1911.01804}
is that now the matter fields directly interact with non-abelian gauge fields.

%\subsection{Degree-2 polynomial symmetry and global-covariant 3-derivative}
\subsection{Degree-(m-1) polynomial symmetry 
to non-abelian higher-rank tensor gauged matter theory
%and global-covariant m-derivative
}

\label{sec:Degree-m-1}

In this subsection, we outline a generalization of previous 
\Sec{sec:Degree-0-Klein-Gordon}
and \Sec{sec:Degree-1-Pretko's}
to a general degree-(m-1) polynomial symmetry
and by gauging it and the particle-hole $\Z_2^C$ symmetry
to obtain a non-abelian higher-rank tensor gauged matter field theory.

\subsubsection{Global symmetry: $\prod_{M=1}^{\tm-1} \U(1)_{x_{n \choose M}^{M}}$}
\label{sec:degree-(m-1)-polynomial-symmetry}

Follow  \Refe{WXY2-1910-1911.01804}, {a degree (m-1)-polynomial symmetry acts on the complex scalar $\Phi(x) \in C$:}
\bea
\Phi\to\re^{\ii Q(x)}\Phi =   \re^{\ii (\Lambda_{i_1,\dots,i_{\tm-1}}  x_{i_1} \dots x_{i_{\tm-1}} +\dots+ \Lambda_{i,j}  x_ix_j + \Lambda_{i}  x_i + \Lambda_0\big)}\Phi.
\eea
Different $\Lambda_{\dots}$ introduce degree U(1) degrees of freedom.
Different U(1) symmetries for different degrees and different $\Lambda_{\dots}$ commute.
We denote such a degree (m-1)-polynomial symmetry structure with several U(1) symmetry 
groups as\footnote{Each of U(1) factors represents a U(1) group. However the product of these U(1) act differently:
the left most U(1) acts globally as 0-form symmetry,
the left second most $ \U(1)_{x_{(n)}}$ acts a vector global symmetry, etc.
Thus we call it a global symmetry structure (instead of a global symmetry group), because
each U(1) has different physical meanings associated to space(/time) coordinates.
Furthermore, we also refer to its gauging as a \emph{gauged structure} (not necessarily the same as the conventional \emph{gauge group}). }
\bea  \label{eq:U1polyG}
 \boxed{\U(1)_{x_{ n \choose {\tm-1}}^{{\tm-1}}} \times
 \dots
 \times \U(1)_{x_{(n)}} \times \U(1)    
= 
{\prod_{M=1}^{\tm-1} \U(1)_{x_{n \choose M}^{M}}}
}.
\eea
The sub-indices of U(1) specifies which $\Lambda_{\dots}$ degree of freedom contributes such a U(1).
Note that  $0 \leq M \leq \tm-1$.
For example, for the degree-1 polynomial symmetry with different $\Lambda_{j}$, we denote
$$
\U(1)_{x_{(n)}}:= \prod_{j=1}^n \U(1)_{x_{j}}, 
$$
{each for different $\Lambda_{j}$}.
In general, we denote
\bea
{\U(1)_{x_{n \choose M}^{M}}:= \prod_{\{j_1,\dots,j_{M}\}} \U(1)_{x_{j_1},\dots,x_{j_{M}}}}, 
\eea
   each for different 
  $\Lambda_{j_1,\dots,j_{M}}$.

\subsubsection{$\Z_2^C$ charge-conjugation (particle-hole) symmetry}

In addition to the polynomial symmetry in \Sec{sec:degree-(m-1)-polynomial-symmetry},
as noticed in \cite{Wang2019aiq1909.13879, WXY2-1910-1911.01804}, we have a
{$\Z_2^C$ charge-conjugation (particle-hole) symmetry}.
It acts on the complex scalar $\Phi$ switching from a particle to an anti-particle.
The $\Z_2^C$ symmetry persists even after we gauge the abelian polynomial-symmetry,
which also acts on the rank-m abelian symmetric tensor $A_{i_1,\cdots,i_{\tm}}$ and
the gauge parameter $\eta_v(x)$ for 
$\Phi \to e^{\ii \eta_v} \Phi$:
\bea
 \Phi &\mapsto&   \Phi^\dagger,\nn\\ 
 A_{i_1,\cdots,i_{\tm}} &\mapsto& - A_{i_1,\cdots,i_{\tm}},\nn \\
 \eta_v(x) &\mapsto& - \eta_v(x).\nn
\eea
The U(1) {degree (m-1)}-polynomial symmetry does not commute with ${\Z_2^C}$ symmetry.
Abbreviate \Eq{eq:U1polyG}'s ${\prod_{M=1}^{\tm-1} \U(1)_{x_{n \choose M}^{M}}}$ as ${{\rm{U}}(1)_{\rm{poly}}}$ symmetry
\bea
U_{\Z_2^C} U_{{\rm{U}}(1)_{\rm{poly}}}  \Phi \;  &=&U_{\Z_2^C} ( e^{\ii Q(x)} \Phi)
=e^{\ii Q(x)} \Phi^\dagger.\nn\\
 U_{{\rm{U}}(1)_{\rm{poly}}} U_{\Z_2^C} \Phi \; &=& U_{{\rm{U}}(1)_{\rm{poly}}} (\Phi^\dagger)
=e^{-\ii Q(x)} \Phi^\dagger.
\eea
So we have indeed a non-abelian/non-commutative global symmetry structure:
\bea
\boxed{{(\prod_{M=1}^{\tm} \U(1)_{x_{n \choose M}^{M}}
 )   \rtimes \Z_2^C} := U_{{\rm{U}}(1)_{\rm{poly}}} \rtimes \Z_2^C }.
\eea

\subsubsection{Non-Abelian/non-commutative gauge structure:  ${\U(1)_{x_{ n \choose {\tm-1}}^{{\tm-1}}}   \rtimes \Z_2^C}$}

Even after we gauge the U(1) polynomial symmetry, we can still observe the gauge transformation of $[U_{{\rm{U}}(1)_{\rm{poly}}}]$
does not commute with the  $\Z_2^C$ global symmetry transformation, which both can act on the $\Phi$ and the rank-m symmetric tensor $A$ respectively:
\bea
U_{\Z_2^C} [U_{{\rm{U}}(1)_{\rm{poly}}}]  \Phi \;  &=&U_{\Z_2^C} ( e^{\ii \eta_v} \Phi)
=e^{\ii \eta_v} \Phi^\dagger.\nn\\ \;
[ U_{{\rm{U}}(1)_{\rm{poly}}}] U_{\Z_2^C} \Phi \; &=&U_{\Z_2^C} (\Phi^\dagger)
=e^{-\ii \eta_v} \Phi^\dagger.\nn\\
U_{\Z_2^C} [U_{{\rm{U}}(1)_{\rm{poly}}}] A_{i_1,\cdots,i_{\tm}}&=&U_{\Z_2^C} ( A_{i_1,\cdots,i_{\tm}}  +\frac{1}{g} \prt_\mu \prt_\nu  \eta_v)
=- A_{i_1,\cdots,i_{\tm}} +\frac{1}{g} \prt_{i_1}  \prt_{i_2}\cdots   \prt_{i_{\tm-1}} \prt_{i_{\tm}}   \eta_v.\nn\\  \;
[ U_{{\rm{U}}(1)_{\rm{poly}}}] U_{\Z_2^C} A_{i_1,\cdots,i_{\tm}}&=&U_{\Z_2^C} ( -A_{i_1,\cdots,i_{\tm}})
=- A_{i_1,\cdots,i_{\tm}} -\frac{1}{g} \prt_{i_1}  \prt_{i_2}\cdots   \prt_{i_{\tm-1}} \prt_{i_{\tm}}   \eta_v.\nn
\eea
By gauging the degree-(m-1) polynomial symmetry and keep \emph{only} the rank-m symmetric tensor $A_{i_1,\cdots,i_{\tm}} $,
we are left with a non-abelian/non-commutative gauge structure
\bea
\boxed{\U(1)_{x_{ n \choose {\tm-1}}^{{\tm-1}}}   \rtimes \Z_2^C}.
\eea

\subsubsection{Non-abelian $[{\U(1)_{x_{ n \choose {\tm-1}}^{{\tm-1}}}   \rtimes \Z_2^C}]$ gauged matter: Polynomial invariant method}
\label{sec:degree-m-1-nab-gauged-M}

We propose a polynomial invariant method to generalize the procedure of
\Sec{sec:degree-1-nab-gauged-M} from a degree-1 polynomial to a generic degree (here a degree-(m-1) polynomial).
First, we determine, 
\bea
D_{i_1}^{c,\Im}  D_{i_2}^{c,\Im} \dots D_{i_{\tm}}^{c,\Im} \log\Phi :=
\frac{P^c_{i_1,\cdots,i_{\tm}}(\Phi,\cdots,(D^{c,\Im})^{\tm}\Phi)}{\Phi^{\tm}}.
\eea
For example, the degree-1 case is obtained in \Eq{eq:DcDclogphi}.
The degree-2 case is obtained in \Refe{WXY2-1910-1911.01804}.\footnote{For a {degree-2 polynomial} symmetry:
$\Phi\to\re^{\ii Q(x)}\Phi =   \re^{\ii ( \Lambda_{i,j}  x_i x_j  + \Lambda_{i}  x_i + \Lambda_0)}\Phi$,
we construct a {covariant 3-derivative (triple-derivative) below.
First, $\log \Phi \to \log\Phi+ \ii Q(x) = \log\Phi+ {\ii ( \Lambda_{i,j}  x_i x_j  + \Lambda_{i}  x_i + \Lambda_0)}$.
we take $\prt_{x_i}\prt_{x_j}\prt_{x_k}:= \prt_{i}\prt_{j}\prt_{k}$ on both sides
$
\prt_{i}\prt_{j}\prt_{k}\log\Phi =
		\frac{P_{i,j,k}(\Phi,\cdots,\prt^3\Phi)}{\Phi^3} \to\prt_{i}\prt_{j}\prt_{k} \log\Phi,
$ which is globally invariant under the {degree-2 polynomial} symmetry:
$$
\frac{P_{i,j,k}(\Phi,\cdots,\prt^3\Phi)}{\Phi^3}=  \frac{{{\Phi^2 (\prt_{i} \prt_{j} \prt_{k} \Phi )
		-3\Phi\big( \prt_{(k} \Phi \prt_{i} \prt_{j)}  \Phi \big)
		  } +  2{ ( \prt_{i} \Phi)( \prt_{j} \Phi)(\prt_{k} \Phi ) } }}{\Phi^3}. 
$$
We use the symmetrized tensor notation:
$T_{(i_1i_2\cdots i_k)} = \frac{1}{k!}\sum_{\sigma\in \rm{S}_k} T_{i_{\sigma 1}i_{\sigma 2}\cdots i_{\sigma k}}$,
with parentheses $(ijk)$ around the indices being symmetrized.
The $\rm{S}_k$ is the symmetric group of $k$ elements.
Since the denominator $\Phi^3 \to \re^{\ii 3Q(x)}\Phi^3$, so does
the numerator ${P_{i,j,k}(\Phi,\cdots,\prt^3\Phi)} \to \re^{\ii 3Q(x)} {P_{i,j,k}(\Phi,\cdots,\prt^3\Phi)} $, which 
we call the denominator and numerator are {3-covariant}.
 Lagrangian thus contains $|{P_{i,j,k}}|^2 := {P_{i,j,k}}(\Phi){P^{i,j,k}}(\Phi^\dagger)$ \cite{WXY2-1910-1911.01804}.}}
The functional $P^c$ should be a generalization of the result obtained in \Refe{WXY2-1910-1911.01804}.
The derivative $(D^{c,\Im})$ involves the coupling to a 1-form $C$ gauge field.
Moreover, when we turn off the $C$ gauge field, we reduce $D_{i_1}^{c,\Im}  D_{i_2}^{c,\Im} \dots D_{i_{\tm}}^{c,\Im} \log\Phi$
to a previous formula obtained in \Refe{WXY2-1910-1911.01804}:
$$
	\prt_{i_1}\cdots\prt_{i_{\tm}}\log\Phi :=\frac{P_{i_1,\cdots,i_{\tm}}(\Phi,\cdots,\prt^{\tm}\Phi)}{\Phi^{\tm}}.
$$
A generic ${\U(1)_{x_{ n \choose {\tm-1}}^{{\tm-1}}}   \rtimes \Z_2^C}$ gauge transformation contains,
 \bea \label{eq:degreem-q-A-gauge-trans}
   A_{\mu \nu}  &\to&   A_{\mu \nu}'= e^{\ii \gamma_c(x)} A_{\mu \nu}  +    (\pm)^{\rtimes} \frac{1}{({\tm}!) g}(D_{(i_1}^c D_{i_2}^c \dots D_{i_{\tm})}^c) ( \eta_v(x)). \quad\quad\quad\\
   \label{eq:degreem-1-C-gauge-trans}
 C_j &\to&  C_{j }'=C_j +\frac{1}{g_c} \prt_j \gamma_c(x). \\
 \label{eq:degreem-1-Phi-gauge-trans}
 \Phi  &\to& \re^{ (\pm)^{\rtimes} \ii \eta(x)} \Phi_c =
  \re^{(\pm)^{\rtimes} \ii \eta(x)} (  \Phi_{\Re} + \ii \re^{\ii \gamma_c(x)} \Phi_{\Im}  )  =
 \left\{\begin{array}{l} 
 \re^{\ii \eta(x)} \Phi , \text{ if } \gamma_c \in  \pi \Z_{\text{even}},\\
 \re^{(\pm)^{\rtimes}\ii \eta(x)} \Phi^\dagger, \text{ if } \gamma_c \in  \pi \Z_{\text{odd}}.  \end{array}\right.\quad\quad
 \eea
 Here $(D_{(i_1}^c D_{i_2}^c \dots D_{i_{\tm})}^c):=
(D_{i_1}^c D_{i_2}^c \dots D_{i_{\tm}}^c
+
D_{i_2}^c D_{i_1}^c \dots D_{i_{\tm}}^c
+\dots)$ yields a symmetrization over the subindices under the lower bracket  $({i_1,\cdots,i_{\tm}})$,
the permutation $({\tm}!)$-terms.
The ${P^c_{i_1,\cdots,i_{\tm}}(\Phi,\cdots,(D^{c,\Im})^{\tm}\Phi)}$ is not gauge covariant under the
generic gauge transformation. But we can append the $A$ gauge field to make it gauge covariant:
\bea  \label{eq:gauge-covariant-D-P-A-C}
        D_{i_1,\cdots,i_{\tm}}^{A,c,\Im}[\{\Phi \}] &:=&{P^c_{i_1,\cdots,i_{\tm}}(\Phi,\cdots,(D^{c,\Im})^{\tm}\Phi)}- \ii g A_{i_1,\cdots,i_{\tm}} \Phi^{\tm},
\eea
where we implicitly sum over all possible indices as
$ \sum_{\{i_1,\cdots,i_{\tm}\}}$ over both the left and right hand sides.
The special case when 
$\tm=1$ is given in \Eq{eq:Gauge-covariant-1-derivative-nab} and
$\tm=2$ is given in \Eq{eq:Gauge-covariant-2-derivative-nab-1}.
The non-abelian gauge covariant rank-(m$+1$) field strength is already obtained and defined in \Refe{WXY2-1910-1911.01804}:
$$
{\hat F^c_{\mu,\nu,i_2,\cdots,i_{\tm}}  :=D_\mu^c A_{\nu,i_2,\cdots,i_{\tm}}  -D_{\nu }^c A_{\mu,i_2,\cdots,i_{\tm}}
:=(\prt_\mu - \ii g_c C_\mu) A_{\nu,i_2,\cdots,i_{\tm}} -(\prt_{\nu} - \ii g_c C_\nu ) A_{\mu,i_2,\cdots,i_{\tm}} }
$$
We can include the ingredients of non-abelian gauge theory coupling to the newly obtained gauged matter sectors 
(see footnote \ref{ft:flip-C})
\bea \label{eq:general-rank-fractonic-gauged-matter}
&&\hspace{-15mm}
\bZ_{\text{rk-(m$+1$)-NAb-$\Phi$}} =\int 
(\prod_{I=1}^{N} [\cD A_I][\cD B_I]  [\cD C_I][\cD \Phi_I][\cD \Phi_I^\dagger])\exp(\ii \int_{M^{d+1}}  \sum_{I=1}^{N} \Big( \dd^{d+1}x\big(  |\hat F^{c,I}_{\mu,\nu,i_2,\cdots,i_{\tm}}|^2 \\
&&\hspace{-15mm}+D_{i_1,\cdots,i_{\tm}}^{A,c,\Im}[\{\Phi_I \}]  D^{\dagger {i_1,\cdots,i_{\tm}}}_{ A,c,\Im}[\{\Phi_I^\dagger \}] 
+D_{i_1,\cdots,i_{\tm}}^{\dagger A,c,\Im}[\{\Phi_I \}]  D^{i_1,\cdots,i_{\tm}}_{A,c,\Im}[\{\Phi_I^\dagger \}] 
 +V(\{ |\Phi_I|^2\})  \big)+ \frac{ 2}{2 \pi} B_I \dd C_I \Big))
\cdot \omega_{d+1}(\{C_I\}),\nn
\eea 
so we derive a
newl non-abelian gauged matter field theory. The setup and notations are directly generalized from \Sec{sec:degree-1-nab-gauged-M}.

\section{New Sigma Models in a Family and Two Types of Vortices}
\label{sec:SigmaModelFamilyandTwoTypesofVortices}

Section \ref{sec:Non-Abelian-Gauged-Fractonic-Matter-Theories} proposes a family of non-abelian gauged matter field theories.
In this section, we study the ``dualized'' theory -- instead of using the matter field degrees of freedom,
we try to incorporate the vortex degrees of freedom into the field theory.

To start with, there are at least two types of vortex degrees of freedom that we can identify.
\begin{enumerate}[leftmargin=2.0mm, label=\textcolor{blue}{\arabic*}., ref=\textcolor{blue}{\arabic*}]
\item \label{vortex-1-phi}
The complex scalar matter field can be written as:
\bea
\Phi (x) &=&\sqrt{\rho (x)} \exp(\ii \phi(x)) \in \bC, \label{eq:Phi-rho-phi}\\
\rho (x) & \in & \R_{\geq 0},\\
\phi (x) & \in & [0, 2 \pi) + 2 \pi \Z.
\eea
So we can 
\Eq{eq:general-rank-fractonic-gauged-matter} replace the path integral measure
$\int [\cD \Phi_I][\cD \Phi_I^\dagger]$ to 
$\int [\cD \rho][\cD \phi]$ (up to some phase space volume factor),
also substitute 
$\Phi_I= \sqrt{\rho_I} \exp(\ii \phi_I) $
and 
$\Phi_I^\dagger= \sqrt{\rho_I} \exp(-\ii \phi_I)$.\footnote{We will capture the
integer non-smooth singular part of
$\frac{1}{2\pi}\dd \dd\phi=n \in \Z$
via 
{Cauchy-Riemann relation, the winding number and topological degree theory}
in
\Sec{sec:Cauchy-Riemann}.
}
The $\R_{\geq 0}$ takes the non-negative real values.
When there are $N$ layers, $I=1,\dots,N$,
there are $N$ flavors of vortex fields $\phi_I$.

\item  \label{vortex-2-theta}
The anti-symmetric tensor TQFT sector (the level-2 BF theory as a $\Z_2$ gauge theory twisted by Dijkgraaf-Witten group cohomology topological terms)
$$
\int \prod_{I=1}^{N}[\cD B_I]  [\cD C_I]
\exp(\ii \int_{M^{d+1}} (\sum_{I=1}^{N} \frac{ 2}{2 \pi} B_I \dd C_I))\cdot \omega_{d+1}(\{C_I\})
$$
can also be regarded as the \emph{disordered phase} of a sigma model
given by another scalar field $\theta_I$. 
The derivations of sigma models governing the ordered-disorder phases relevant for twisted Dijkgraaf-Witten type TQFTs
\footnote{\label{footnote:vortex}
To recall the approach of \Refe{Gu2015lfa1503.01768, YeGu2015eba1508.05689} and their generalization: 
\begin{itemize}
\item
The ordered phase of sigma models describes the weak fluctuations around the symmetry-breaking phases.
\item The disorder phases of sigma models
describes the strong fluctuations around the symmetry-restored phases as continuum formulations of TQFTs, SETs or SPTs of Dijkgraaf-Witten type.
\end{itemize}
{{
The U(1) spontaneously symmetry breaking phase has a superfluid ground state, which is an ordered phase respect to $\phi$  with an order parameter 
$$\langle \exp(\ii \theta) \rangle\neq 0.$$
It is well-known that if we disorder the U(1) spontaneously symmetry breaking (superfluid) state, %by \cred{condensing the quantum phase vortices},
we can obtain an disordered phase known as a gapped insulator \cite{{Fisher-Lee},{Dasgupta-Halperin},{Nelson}}.
Our approach is basically along this logical thinking, except that we generalize the approach by: 
\begin{itemize}
\item Disordering the ordered phase (U(1) symmetry breaking superfluid) to a disordered phase of gapped topological order (e.g. the $\Z_{\rN}$-gauge theory, where
the $\Z_{1}$-gauge theory means a trivial gapped insulator,
and the $\Z_{2}$-gauge theory means the low energy theory of deconfined $\Z_{2}$-toric code, $\Z_{2}$-spin liquid or $\Z_{2}$-superconductor).
(Beware that  \Refe{Gu2015lfa1503.01768, YeGu2015eba1508.05689} only consider the case of a superfluid-insulator transition for $\rN=1$, 
here we consider a superfluid-topological-order transition for a generic N.
\item Follow \cite{Gu2015lfa1503.01768, YeGu2015eba1508.05689}, introducing additional topological multi-kink Berry phase specified by the cocycle of cohomology group 
$$\omega_{d+1}(\{C_I\}) \simeq \omega_{d+1}(\{\dd \theta_I\})$$ 
 to the superfluid.
\end{itemize}
To comprehend our formalism, here we overview this approach using field theory \cite{Gu2015lfa1503.01768}.
We start from the superfluid state in a $d$-spacetime dimension described by a bosonic U(1) quantum phase $\theta$ kinetic term
and a superfluid compressibility coefficient $\chi$, the partition function $\mathbf{Z}$ is:
\begin{eqnarray} %(\partial_\mu \theta_{\text{smooth}}+\partial_\mu \theta_{\text{singular}})
\mathbf{Z}=\int [\cD \theta] \exp( - \int \dd^d x \, \frac{\chi}{2} (\partial_\mu \theta_{\text{s}}+\partial_\mu \theta_{\text{v}})^2 ).
\end{eqnarray}
The $\theta = \theta_{\text{s}}+\theta_{\text{v}}$ with a smooth piece $\theta_{\text{s}}$ and a singular vortex piece $\theta_{\text{v}}$ for the bosonic phase $\theta$.
We emphasize that the $\theta_{\text{v}}$ is essential to capture the vortex core, see \Sec{sec:Cauchy-Riemann}.
We introduce an auxiliary field $j^\mu$ and apply the Hubbard-Stratonovich technique \cite{1959PhRvLHubbard},
%\begin{eqnarray} 
\begin{equation}
\mathbf{Z}=\int [\cD \theta] [\cD j^\mu] \exp( - \int  \dd^d x \,
\frac{1}{2\chi}(j^\mu_I)^2 - \ii  j^\mu (\partial_\mu \theta_{\text{s}}+\partial_\mu \theta_{\text{v}})).
\end{equation}
%\end{eqnarray}
By integrating out the smooth part $\int [D \theta_{\text{s}}]$, we obtain a constraint $\delta(\partial_\mu j^\mu)$ into the path integral measure.
Naively, in the anti-symmetric tensor differential form notation, 
 the constraint seems in disguise
$$
\dd (\star j)=0,  \quad \text{ or, }  \quad 2 \pi  \oint \dd (\star j)=\Z,
$$
and the 
solution in disguise is  $j= \frac{1}{2\pi} (\star \dd B).$ (Here we choose a normalization convention.)
However, we imagine the procedure is the N\emph{-fold vortex of superfluid becomes a trivial object} (\emph{instead of a 1-fold vortex of superfluid}) 
that can be created or annihilated for free from the $\Z_{\rN}$-gauge theory vacuum.
Instead we may impose a revised constraint  
\bea
2 \pi \oint \dd (\star j)= \rN \Z. 
\eea
Note that the $2 \pi \rN$ on the right hand side means that N-fold of $2 \pi$ vortices become to be identified as a trivial zero vortex (none vortex).
The solution is, with $\star$ the Hodge star,
\bea \label{eq:j-dB}
j= \frac{\rN}{2\pi} (\star \dd B).
\eea 
We can define a generic form 
$
j^\mu=\frac{\rN}{2 \pi (d-2)!} \epsilon^{\mu \mu_2  \dots \mu_{d} } \partial_{\mu_2} B_{\mu_3 \dots \mu_{d}},
$
with an anti-symmetric tensor $B$ with a total spacetime dimension $d$ (most conveniently,
we may consider 2d space or 2+1d spacetime in order to implement a winding number in \Sec{sec:Cauchy-Riemann}), to satisfy this constraint.}
To disorder the superfluid, we have to make the $\theta_{\text{v}}$-angle strongly fluctuates ---
namely we should take the $\chi \to \infty$ limit  to achieve large $|\delta \theta_{\text{v}}|^2 \gg 1$, the disordered limit of superfluid. 
Plug in \Eq{eq:j-dB}, the partition function becomes:
$$
\mathbf{Z}=\int [\cD \theta_{\text{v}}] [\cD B] \exp( + \int 
 \ii  \frac{\rN}{2\pi}   B  \wedge  (\dd^2 \theta_{\text{v}})  ).
 $$
Hereafter we may compensate the dropped $\pm$-sign by a field-redefinition. 
Although naively $\dd^2=0$, due to the singularity core of $\theta_{\text{v}}$, 
the $(\dd^2 \theta_{\text{v}})$ can be nonzero, see \Sec{sec:Cauchy-Riemann},
which implies that (at least for the 2-dimensional space mapping to a deformed $S^1$-circle as a target space):
\bea
\frac{1}{2\pi}\dd^2 \theta_{\text{v}} =n \mod \rN, \text{ thus } n \in \Z_{\rN}.
\eea
Thus, $(\dd^2 \theta_{\text{v}})$ describes the vortex core density and the vortex current, which we denote 
$$
\frac{1}{2\pi}\dd^2 \theta_{\text{v}} =\star j_{\text{vortex}}.
$$
In addition, Noether theorem leads to the conservation of the vortex current: the continuity equation
$$
\dd \star j_{\text{vortex}}=0,
$$ 
this implies that 
$$
\star j_{\text{vortex}}=\dd C/(2\pi)
$$ for some 1-form gauge field $C$.
We can thus define the singular part of bosonic phase 
$$
\dd \theta_{\text{v}}=C
$$ 
as a 1-form gauge field, to describe the vortex core.
The partition function in the disordered state away from the superfluid, now becomes
that of an gapped insulator (for N = 1) or topologically ordered state with a topological level-N BF action as a $\Z_{\rN}$-gauge theory:
\begin{equation} \label{ed:Z-BF}
\mathbf{Z}=
\int [\cD b] [\cD a] \exp( \frac{\ii}{2\pi} \int  B \wedge \dd C \,)=
\int [\cD b] [\cD a] \exp( \ii \int  
\frac{\dd^d x}{2 \pi (d-2)!} \epsilon^{\mu \mu_2  \dots \mu_{d} } B_{\mu \mu_2 \mu_3 \dots} \partial_{\mu_{d-1}}   C_{\mu_{d}}
).
\end{equation}
}
} 
had been studied in \cite{Ye2014oua1410.2594PengYe, Gu2015lfa1503.01768, YeGu2015eba1508.05689},
here we will implement the procedure done in  \cite{Gu2015lfa1503.01768}.
We write
\bea
\theta_I= \theta_{s,I} +  \theta_{v,I},
\eea
where
$\theta_{s,I}$ describes the smooth (s) part while the
$\theta_{v,I}$ describes the singular vortex (v) part.
See the footnote \ref{footnote:vortex} and \cite{Gu2015lfa1503.01768}, the exterior derivative of the vortex field should be identified
as the 1-form $C$ gauge field as: 
\bea \label{eq:thetavC}
\dd \theta_{\text{v}}=C.
\eea
In this way, the TQFT sector of the theory (as a disordered phase of some sigma model) can be re-written as
a sigma model with the vortex field $\theta_{\text{v}}$ degree of freedom:
\begin{multline} \label{eq:2}
\int (\prod_{I=1}^{N} [\cD B_I]  [\cD \theta_{s,I}][\cD \theta_{v,I}])\exp(\ii \int \limits_{M^{d+1}} \Big( \ii
(
 \sum_{I=1}^{N} {\chi_I\over2}  ( \dd \theta_{s,I} + \dd \theta_{v,I}) \wedge \star( \dd \theta_{s,I} + \dd \theta_{v,I})\\
+
 \frac{ 2}{2 \pi} \sum_{I=1}^{N} B_I \dd (\dd \theta_{v,I}))\Big) \cdot \omega_{d+1}(\{\dd \theta_{v,I}\}).
 \end{multline}
 where
 $ \omega_{d+1}(\{\dd \theta_{v,I}\})$ is mapped to a multi-kink Berry phase topological term
 $\exp(\ii \int \limits_{M^{d+1}}   \# (\dd \theta_{v,1})\wedge \dots \wedge (\dd \theta_{v,N}))$ \cite{Gu2015lfa1503.01768, YeGu2015eba1508.05689}.
\end{enumerate}
We replace and redefine a new derivative 
on the right hand side by substituting $C=\dd \theta_{\text{v}}$ (or $C_{\mu}=\prt_{\mu} \theta_{\text{v}}$):\footnote{
We also have the gauge transformation descended from 1-form $C$ gauge field,
\bea
 \prt_\nu \theta_{v,I} &\to&  \prt_\nu \theta_{v,I} +\frac{1}{g_c} \prt_\nu \gamma_{c,I}(x),\nn \\
 \theta_{v,I} &\to& \theta_{v,I} +\frac{1}{g_c} \gamma_{c,I}(x).\nn
\eea}
\bea
D_{\mu}^{c,\Im} \Phi \to D_{\mu}^{\dd \theta_{\text{v}},\Im} \Phi
&:=&
\prt_{\mu} \Phi -\ii g_c (\prt_{\mu} \theta_{\text{v}}) \Phi_{\Im},\\
D_{\mu}^{A,c,\Im} \Phi 
\to D_{\mu}^{A,\dd \theta_{\text{v}},\Im} \Phi
&:=&
D_{\mu}^A \Phi -\ii g_c (\prt_{\mu} \theta_{\text{v}})
\Phi_{\Im}=(\prt_{\mu} - \ii g A_{\mu})\Phi -\ii g_c (\prt_{\mu} \theta_{\text{v}}) \Phi_{\Im}. \label{eq:Gauge-covariant-1-derivative-nab} \\
 D_\mu^c A_{\nu,i_2,\cdots,i_{\tm}} \to D_\mu^{\dd \theta_{\text{v}}} A_{\nu,i_2,\cdots,i_{\tm}}
&:=&(\prt_\mu - \ii g_c (\prt_{\mu} \theta_{\text{v}})) A_{\nu,i_2,\cdots,i_{\tm}} .\\
\hat F^c_{\mu,\nu,i_2,\cdots,i_{\tm}}  \to \hat F^{{\dd \theta_{\text{v}}}}_{\mu,\nu,i_2,\cdots,i_{\tm}}  &:=&D_\mu^{{\dd \theta_{\text{v}}}} A_{\nu,i_2,\cdots,i_{\tm}}  -D_{\nu }^{{\dd \theta_{\text{v}}}} A_{\mu,i_2,\cdots,i_{\tm}}.
\eea
We can either include or omit the $I$ index for these operators. 

%
%{\begin{center}\fbox{\parbox{7in}{\parindent=0pt 
%}}\end{center}}

By combining two kinds of vortex degrees of freedom from
the vortex \ref{vortex-1-phi} of $\phi$ and the vortex \ref{vortex-2-theta} of $\theta$,
thus we can rewrite \Eq{eq:general-rank-fractonic-gauged-matter} into a sigma model-like expression for a non-abelian gauged fractonic matter theory:
\bea 
&&\hspace{-15mm} \label{eq:general-rank-fractonic-gauged-matter-Sigma}
\bZ_{\text{rk-(m$+1$)-NAb-\text{vortex}}}^{\text{Sigma model}} =\int 
(\prod_{I=1}^{N} [\cD A_I][\cD B_I]   [\cD \theta_{s,I}][\cD \theta_{v,I}][\cD \rho_I][\cD \phi_I])\nn\\
&&\hspace{-15mm}
\exp(\ii \int_{M^{d+1}}  \sum_{I=1}^{N} \Big( \dd^{d+1}x\big(  |\hat F^{\dd \theta_{v},I}_{\mu,\nu,i_2,\cdots,i_{\tm}}|^2 
+
 \sum_{I=1}^{N} {\chi_I\over2}  | \dd \theta_{s,I} + \dd \theta_{v,I}|^2\nn\\
&&\hspace{-15mm}+D_{i_1,\cdots,i_{\tm}}^{A,\dd \theta_{v} ,\Im}[\{\sqrt{\rho_I} \exp(\ii \phi_I)  \}]  D^{\dagger {i_1,\cdots,i_{\tm}}}_{ A,\dd \theta_{v} ,\Im}[\{ \sqrt{\rho_I} \exp(-\ii \phi_I)   \}] 
%\nn\\
%&&\hspace{-15mm}
+D_{i_1,\cdots,i_{\tm}}^{\dagger A,\dd \theta_{v} ,\Im}[\{\sqrt{\rho_I} \exp(\ii \phi_I)  \}]  D^{i_1,\cdots,i_{\tm}}_{A,\dd \theta_{v} ,\Im}[\{\sqrt{\rho_I} \exp(-\ii \phi_I)  \}] 
\nn\\
&&\hspace{-15mm} 
 +V(\{ |\rho_I|\})  \big)
+ \frac{ 2}{2 \pi} B_I \dd (\dd \theta_{v,I}) \Big))
\cdot \omega_{d+1}(\{\dd \theta_{v,I}\}).
\eea 
Here we have substituted \Eq{eq:Phi-rho-phi}, \Eq{eq:thetavC} and $ |\Phi_I|^2=|\rho_I|$.
This \Eq{eq:general-rank-fractonic-gauged-matter-Sigma}
is the most generic form of sigma model -- a part of its phase diagram gives rise to the TQFT (when the $\theta_{v,I}$ vortices disordered),
while the other  part of its phase diagram gives rise to the spontaneously symmetry breaking superfluid like phases.
We emphasize that
$\dd (\dd \theta_{v,I}) =  \dd^2  \theta_{v,I}$ is nonzero and can be
related to a quantized number such as a winding number at the core of the vortex field, see \Sec{sec:Cauchy-Riemann}.

Here $B$ is only a Lagrange multiplier. Here we also have not yet replaced the symmetric tensor gauge field $A$ to any kinds of vortex degrees of freedom in
\Eq{eq:general-rank-fractonic-gauged-matter-Sigma}. As some of the readers may wonder, and it is tempting to ask this question: whether the gauge field 
$A$ can be ``dualized'' into some new vortex degrees of freedom.
However, we will not attempt to attack this issue and leave this as an open question for future work.

\section{Cauchy-Riemann Relation, Winding Number, and Topological Degree Theory}
\label{sec:Cauchy-Riemann}

Here we derive a relation used in the previous section, relating 
the vortex degrees of freedom to a winding number, via the  
Cauchy-Riemann relation and topological degree theory, at least for the 2-dimensional space mapping to a deformed $S^1$-circle as a target space.
On the complex plane 
\bea
z := x+ \ii y = r \exp(\ii \varphi),
\eea 
we define the Hodge star operator $\star$ (for the differential form, this is the Hodge dual) on the 1-form as 
\bea
\star (f\dd x+g\dd y)=(-g\dd x+f\dd y),
\eea
for some generic functions $f$ and $g$.

Then we have $\star \dd z=-\ii \dd z$ with $\dd z=\dd x+\ii \dd y $. We can compute that 
\bea
\dd \star \dd f=(\Delta f ) (\dd x\wedge \dd y),
\eea
with the Laplacian or Laplace operator $\Delta \equiv \nabla^2$.

If $f=u+\ii v$ is holomorphic dependent on $z$ independent of $\bar{z}$ (namely the Cauchy-Riemann equation), 
$\dd f=f' \dd z$ and also $\ii \star \dd f=f' \dd z$ with $f'= \frac{\dd f}{\dd z}$, so we see that $\dd v=\star \dd u$.

Take $f=\log z =  \log (r \exp(\ii \varphi) )= \log  r +\ii \varphi$, then $u=\log r$ and $v=\varphi$, where $z=r \re^{\ii \varphi}$ is the polar coordinate, then we have 
\bea
\dd \dd \varphi= \dd \dd v=\dd \star \dd u=(\Delta \log r ) (\dd x\wedge \dd y)= 2\pi \delta_0
\eea
where 
$$
\delta_0 \equiv \delta(\mathbf{r}) (\dd x\wedge \dd y)
$$ is the delta function at the origin $0$ of the polar coordinate.\footnote{Here we use the facts about the
fundamental solutions of Laplace's or Poisson's equations by solving the source of Dirac delta function $\delta(r)$, see for example \cite{evans10}.
The volume of $D$-dimensional ball $B^D$ is $V_D \equiv \frac{\pi^{D/2}}{\Gamma({D/2+1})}$, where
the gamma function obeys
$\Gamma(n+1)=n \Gamma(n)$ for general $n \in \mathbb{C}$, ${\Gamma({1/2})}=\sqrt{\pi}$,  while $\Gamma(n)=(n-1)!$ for any positive integer $n$. The area of the boundary of 
the $D$-dimensional ball $B^D$ is the hypervolume of the $(D-1)$-dimensional sphere, denoted as $A_{B^D} \equiv A_{S^{D-1}} = D V_D \equiv \frac{D\pi^{D/2}}{\Gamma({D/2+1})}
=\frac{2\pi^{D/2}}{\Gamma({D/2})}$. Given the vector $\mathbf{r} \in \mathbb{R}^D$ and the radial $r= |\mathbf{r} |$,
then we have the following properties for Laplacian $\Delta = \nabla^2$:
\bea
\begin{cases}
D=2, \quad\nabla^2\bigg(\ln(r)\bigg)=2\pi\delta^{(2)}(\mathbf{r}).\\
D=3, \quad \nabla^2\bigg( r^{-1}\bigg)=-4\pi\delta^{(3)}(\mathbf{r}).\\
\dots\\
D \geq 3, \quad \nabla^2\bigg( r^{-(D-2)}\bigg)=- (D-2)A_{S^{D-1}} \delta^{(D)}(\mathbf{r}).
\end{cases}
\eea
In general $\delta_0 \equiv \delta(\mathbf{r}) (\text{vol}_D) = \delta(\mathbf{r}) (\dd x_1 \wedge \dots \wedge \dd x_D) $.
}
Hence we derive that 
\bea
\frac{1}{2\pi}\dd \dd \varphi=\delta_0.
\eea
For a general $S^1$ valued function $\phi$ defined outside a singular point $p$, we may assume $p=0$ at the origin (without loss of generality). 

Applications to two types of vortices:\\[-10mm]
\begin{enumerate}[leftmargin=2.0mm, label=\textcolor{blue}{\arabic*}., ref=\textcolor{blue}{\arabic*}]
\item
For the fractonic phase field $\phi$:
We can always write 
the phase of the fractonic matter field in \Eq{eq:Phi-rho-phi} as
\bea
\phi :=\phi_{s} + \phi_{v}: =\phi_{s} + n \varphi
\eea
where $\phi_{s}$ captures the smooth ($s$) part and 
$\phi_{v}$ captures the singular vortex ($v$) part, while $n$ is the winding number. In terms of the degree theory, we only focus on
$\Sigma^2 \to S^1$, specifically here we consider $\Sigma^2=\mathbb{R}^2 -\{0\}$, as a punctured 2-plane mapping to a circle $S^1$ (or the U(1) target space).
 We can possibly generalize this result to other target spaces. 
The first term $\phi_{s}$ extends smoothly over $0$ is known as the smooth fluctuation. 
Then using the previous result we see that 
\bea
\frac{1}{2\pi}\dd \dd\phi=\frac{1}{2\pi}\dd \dd (\phi_{s} + \phi_{v}) = \frac{1}{2\pi}\dd \dd( \phi_{v})=n \frac{1}{2\pi}\dd \dd(   \varphi)=n \; \delta_0, \text{ with } n \in \Z. 
\eea
Thus importantly, we can identify the solution of vortex core equation $ \frac{1}{2\pi}\dd^2 \phi_{v}= n  \;\delta_0$ with $n \in \Z$
as the winding number. %of the vortex field
%%%
\item 
For the $\Z_2^C$-gauge phase field $\theta$:
Similarly, for the vortex associated to the $C$ gauge field,
\bea
\frac{1}{2\pi}\dd \dd\theta=\frac{1}{2\pi}\dd \dd (\theta_{s} +\theta_{v}) = \frac{1}{2\pi}\dd \dd( \theta_{v})=n_{\theta_{v}} 
\; \delta_0, \text{ with } n_{\theta_{v}} \in \Z. 
\eea
So if the Cauchy-Riemann relation can be applied, we can rewrite:
$$
 \frac{ 2}{2 \pi} B_I \dd (\dd \theta_{v,I}) =  B_I \; n_{\theta_{v}} \; \delta_0$$
to a term associated to the winding number $n_{\theta_{v}}$.
We can adjust 
$\delta_0$ from the location of the origin to other points $p$ as
$\delta_p$ for all the above discussions to indicate the locations of vortices at points $p$.

\end{enumerate}

\section{Conclusions}
\label{sec:Conclusions}

We have proposed a systematic framework to obtain a family of {non-abelian gauged fractonic matter field theories} in \Sec{sec:Non-Abelian-Gauged-Fractonic-Matter-Theories}.
We have derived a new family of Sigma models with two types of vortices in \Sec{sec:SigmaModelFamilyandTwoTypesofVortices}
that can interplay and transient between the disordered phase (with higher-rank tensor non-abelian gauge theories coupled to fractonic matter)
and ordered phase (with superfluids and vortex excitations) of Sigma models.
We formulate two types of vortices, one is associated to the fractonic matter fields ($\dd (\dd \phi)$),
the other is associated to the 1-form $C$ gauge field ($\dd (\dd \theta_{v})$). 
These two types of vortices mutually interact non-commutatively when they communicate via the higher-rank tensor gauge field $A$ as a propagator.
We apply the
{Cauchy-Riemann relation and topological degree theory to capture the winding number}
for the two types of vortices in \Sec{sec:Cauchy-Riemann}.

Here we make some extended comments and also list down some {open questions}.
\\[-10mm]
\begin{enumerate}[leftmargin=2.0mm, label=\textcolor{blue}{\arabic*}., ref=\textcolor{blue}{\arabic*}]

\item \emph{Reduction to the abelian case}: When the 1-form gauge field $C=0$,
we reduce to the abelian tensor gauged matter theory. If we further turn off the tensor gauge field $A$, then our sigma model should reduce to a simplified
special case of the abelian fractonic superfluid models in \Refe{Yuan2019gehPengYe1911.02876}.
When we turn on the gauge field $C$, there are nontrivial couplings between matter sectors and anti-matter sectors,
because the particle-hole symmetries are dynamically gauged.

\item Our Sigma model \Eq{eq:general-rank-fractonic-gauged-matter-Sigma} contains the target space with radius size $\sqrt \rho$.

If the $\sqrt \rho$ is fixed, then we have a fixed radius $S^1$ target space, we only have the fluctuation around the $\phi$ fields (e.g. Goldstone modes, superfluid, or vortices).

If the $\sqrt \rho$ also fluctuates, the volume and radius size of Sigma model target space change, 
so we have more interesting dynamics for the Sigma model. Its dynamics and low energy fates are interesting, but
perplexing and challenging, which are important questions for the future.

\item We had mentioned \emph{two types} of vortices mutually interact via tensor gauge field $A$, causing some kind of non-abelian vortex behavior altogether. 
Moreover, it is tempting to know whether we can also dualize tensor gauge field $A$ to represent the third type of vortex from $A$.

\item Our formulation of Sigma models may have applications to superfluid, supersolid, quantum melting transition, and elasticity studied in the recent fracton literature (for selected references, quantum crystal disclinations and dislocations,
see\cite{2018PRLPretkoLeo1711.11044, 2018PRLPretkoLeo1808.05616, 2019LeoHermelearXiv1905.06951, 2019PretkoarXivCrystal-to-Fracton-1907.12577, Gromov2019waaPiotr1908.06984, YouPretko2019cvs1908.08540plaquette-melting-transitions} and citations therein).

\item In the present literature, there are \emph{three different routes} to obtain non-abelian fracton orders:\\
(1) Gauge the charge conjugation (i.e., particle-hole) $\Z_2^C$-symmetry and ${{\rm{U}}(1)_{\rm{poly}}}$ polynomial symmetry \cite{Wang2019aiq1909.13879, WXY2-1910-1911.01804},\\
(2) Gauge the permutation symmetry of $N$-layer systems \cite{BulmashMaissamBarkeshli2019taq1905.05771, PremWilliamson2019etl1905.06309},\\
(3) Couple to non-abelian TQFT/topological order \cite{Vijay2017cti1706.07070, SongPremHuang2018gbb1805.06899,  PremHuangSong2018jsn1806.04687} and  \cite{Wang2019aiq1909.13879, WXY2-1910-1911.01804}.\\
Given a $N$-layer systems, there is a larger non-abelian group structure that we can explore.
In the previous work \cite{Wang2019aiq1909.13879, WXY2-1910-1911.01804} and our present work,
we focus on the finite abelian group $(\Z_2^C)^N$ by gauging $N$-layer of particle-hole symmetries,
and consider a non-abelian gauge structure: ${{\rm{U}}(1)_{\rm{poly}}} \rtimes (\Z_2^C)^N$.
In fact, a natural larger group is including also the $\rm{S}_N$ permutation symmetry group of $N$-layers  \cite{BulmashMaissamBarkeshli2019taq1905.05771, PremWilliamson2019etl1905.06309}.
%,which gives rise to an additional  $\rm{S}_N$ permutation symmetry.
The $(\Z_2^C)^N$ and $\rm{S}_N$ form a short exact sequence via a group extension:
\bea
1\to (\Z_2^C)^N \to G_{\text{nAb}} \to \rm{S}_N \to 1.
\eea
The $G_{\text{nAb}}$ is related to the \emph{hyperoctahedral group} in mathematics.
An overall larger non-abelian group structure including both $G_{\text{nAb}}$ and ${{\rm{U}}(1)_{\rm{poly}}}$,
 that mutually are non-commutative, possibly can be studied also via field theories or lattice models in the future.

\item \Refe{WXY2-1910-1911.01804} points out possible proper tools for studying these field theories include algebraic variety and affine-geometry/manifold.
One motivation is to explore these theories and more general models on general affine manifolds beyond the Euclidean spacetime.
This is left for future work.

\item Quantization and quantum path integral: 
Although we write down schematic path integral forms, our analysis is mostly based on equations of motion (EOM) and semi-classical
or mean-field analyses. 
The most challenging question may be
to explore the full quantum nature of the path integral we proposed, or study the 
quantization of these field theories. However, 
since we do have a \emph{quantum mechanical definition} of the theory on a regularized lattice with an energy cutoff:  
\begin{itemize}
\item the U(1) tensor field theory on the lattice \cite{2016arXiv160108235RRasmussenYouXu} \cite{RahulNandkishore2018sel1803.11196, 2001.01722PretkoChenYou}, and
\item the topological gauge theory from group cohomology (of Dijkgraaf-Witten type) on the lattice \cite{Wan1211.3695, Wan2014woaWWH1409.3216, Song2018gbb1805.06899},
\end{itemize}
so we believe that our field theories (as the interplay between the two models) should be a promising quantum theory, and \emph{quantum field theories} can be made to be 
mathematically rigorously well-defined.
\end{enumerate}

\section{Acknowledgements} %Acknowledgements

JW thanks Kai Xu and John Loftin for enlightening conversations, and for a collaboration on an upcoming work.
%The authorship is listed in the alphabetical order by the standard convention. 
The method of deriving a family of sigma model theories were partially originated from \cite{Gu2015lfa1503.01768} and an unpublished work of 2015-2016.
%%%%%%%%%%%%%%%%%%%%%
JW thanks the feedback from Abhinav Prem and Yunqin Zheng, %on the manuscript,
and the attendees for Quantum Matter in Math and Physics workshop \cite{JWangCMSA2019-QM}. 
JW was supported by
NSF Grant PHY-1606531 and Institute for Advanced Study. 
% 
%KX  is supported by Harvard Math Graduate Program and
%``The Black Hole Initiative: Towards a Center for Interdisciplinary Research,'' Templeton Foundation. 
This work is also supported by 
NSF Grant DMS-1607871 ``Analysis, Geometry and Mathematical Physics'' 
and Center for Mathematical Sciences and Applications at Harvard University.

\bibliographystyle{Yang-Mills.bst}
\bibliography{fracton-embeddon-sigma.bib}

\providecommand{\href}[2]{#2}\begingroup\raggedright\begin{thebibliography}{10}

\bibitem{RahulNandkishore2018sel1803.11196}
R.~M. Nandkishore and M.~Hermele, \emph{{Fractons}},
  \href{http://dx.doi.org/10.1146/annurev-conmatphys-031218-013604}{\emph{Ann.
  Rev. Condensed Matter Phys.} {\bf 10} 295--313 (2019)},
  [\href{https://arxiv.org/abs/1803.11196}{{\tt arXiv:1803.11196}}].

\bibitem{Pretko2018jbi1807.11479}
M.~Pretko, \emph{{The Fracton Gauge Principle}},
  \href{http://dx.doi.org/10.1103/PhysRevB.98.115134}{\emph{Phys. Rev.} {\bf
  B98} 115134 (2018)}, [\href{https://arxiv.org/abs/1807.11479}{{\tt
  arXiv:1807.11479}}].

\bibitem{SeibergF2019}
N.~Seiberg, \emph{{Field Theories With a Vector Global Symmetry}},
  \href{https://arxiv.org/abs/1909.10544}{{\tt arXiv:1909.10544}}.

\bibitem{Wang2019aiq1909.13879}
J.~Wang and K.~Xu, \emph{{Higher-Rank Tensor Field Theory of Non-Abelian
  Fracton and Embeddon}},  \href{https://arxiv.org/abs/1909.13879}{{\tt
  arXiv:1909.13879}}.

\bibitem{Gromov2018nbv1812.05104}
A.~Gromov, \emph{{Towards classification of Fracton phases: the multipole
  algebra}}, \href{http://dx.doi.org/10.1103/PhysRevX.9.031035}{\emph{Phys.
  Rev.} {\bf X9} 031035 (2019)}, [\href{https://arxiv.org/abs/1812.05104}{{\tt
  arXiv:1812.05104}}].

\bibitem{WXY2-1910-1911.01804}
J.~Wang, K.~Xu and S.-T. Yau, \emph{{Higher-Rank Tensor Non-Abelian Field
  Theory: Higher-Moment or Subdimensional Polynomial Global Symmetry, Algebraic
  Variety, Noether's Theorem, and Gauge}},
  \href{https://arxiv.org/abs/1911.01804}{{\tt arXiv:1911.01804}}.

\bibitem{GriffinPetrHorava2014bta1412.1046}
T.~Griffin, K.~T. Grosvenor, P.~Horava and Z.~Yan, \emph{{Scalar Field Theories
  with Polynomial Shift Symmetries}},
  \href{http://dx.doi.org/10.1007/s00220-015-2461-2}{\emph{Commun. Math. Phys.}
  {\bf 340} 985--1048 (2015)}, [\href{https://arxiv.org/abs/1412.1046}{{\tt
  arXiv:1412.1046}}].

\bibitem{2005PRL0404182Chamon}
C.~{Chamon}, \emph{{Quantum Glassiness in Strongly Correlated Clean Systems: An
  Example of Topological Overprotection}},
  \href{http://dx.doi.org/10.1103/PhysRevLett.94.040402}{\emph{"Phys. Rev.
  Lett."} {\bf 94} 040402 (2005 Jan)},
  [\href{https://arxiv.org/abs/cond-mat/0404182}{{\tt
  arXiv:cond-mat/0404182}}].

\bibitem{1108.2051CastelnovoChamon}
C.~{Castelnovo} and C.~{Chamon}, \emph{{Topological quantum glassiness}},
  \href{http://dx.doi.org/10.1080/14786435.2011.609152}{\emph{Philosophical
  Magazine} {\bf 92} 304--323 (2012 Jan)},
  [\href{https://arxiv.org/abs/1108.2051}{{\tt arXiv:1108.2051}}].

\bibitem{1926PhRvSchrodinger}
E.~{Schr{\"o}dinger}, \emph{{An Undulatory Theory of the Mechanics of Atoms and
  Molecules}}, \href{http://dx.doi.org/10.1103/PhysRev.28.1049}{\emph{Physical
  Review} {\bf 28} 1049--1070 (1926 Dec)}.

\bibitem{1926Klein}
O.~{Klein}, \emph{{Quantentheorie und f{\"u}nfdimensionale
  Relativit{\"a}tstheorie}},
  \href{http://dx.doi.org/10.1007/BF01397481}{\emph{Zeitschrift fur Physik}
  {\bf 37} 895--906 (1926 Dec)}.

\bibitem{1926Gordon}
W.~{Gordon}, \emph{{Der Comptoneffekt nach der Schr{\"o}dingerschen Theorie}},
  \href{http://dx.doi.org/10.1007/BF01390840}{\emph{Zeitschrift fur Physik}
  {\bf 40} 117--133 (1926 Jan)}.

\bibitem{2016arXiv160108235RRasmussenYouXu}
A.~{Rasmussen}, Y.-Z. {You} and C.~{Xu}, \emph{{Stable Gapless Bose Liquid
  Phases without any Symmetry}}, {\emph{arXiv e-prints} arXiv:1601.08235 (2016
  Jan)}, [\href{https://arxiv.org/abs/1601.08235}{{\tt arXiv:1601.08235}}].

\bibitem{Pretko2016kxt1604.05329}
M.~Pretko, \emph{{Subdimensional Particle Structure of Higher Rank U(1) Spin
  Liquids}}, \href{http://dx.doi.org/10.1103/PhysRevB.95.115139}{\emph{Phys.
  Rev.} {\bf B95} 115139 (2017)}, [\href{https://arxiv.org/abs/1604.05329}{{\tt
  arXiv:1604.05329}}].

\bibitem{Pretko2016lgv1606.08857}
M.~Pretko, \emph{{Generalized Electromagnetism of Subdimensional Particles: A
  Spin Liquid Story}},
  \href{http://dx.doi.org/10.1103/PhysRevB.96.035119}{\emph{Phys. Rev.} {\bf
  B96} 035119 (2017)}, [\href{https://arxiv.org/abs/1606.08857}{{\tt
  arXiv:1606.08857}}].

\bibitem{Pretko2017xar1707.03838}
M.~Pretko, \emph{{Higher-Spin Witten Effect and Two-Dimensional Fracton
  Phases}}, \href{http://dx.doi.org/10.1103/PhysRevB.96.125151}{\emph{Phys.
  Rev.} {\bf B96} 125151 (2017)}, [\href{https://arxiv.org/abs/1707.03838}{{\tt
  arXiv:1707.03838}}].

\bibitem{Slagle2018kqf1807.00827}
K.~Slagle, A.~Prem and M.~Pretko, \emph{{Symmetric Tensor Gauge Theories on
  Curved Spaces}},
  \href{http://dx.doi.org/10.1016/j.aop.2019.167910}{\emph{Annals Phys.} {\bf
  410} 167910 (2019)}, [\href{https://arxiv.org/abs/1807.00827}{{\tt
  arXiv:1807.00827}}].

\bibitem{PremHuangSong2018jsn1806.04687}
A.~Prem, S.-J. Huang, H.~Song and M.~Hermele, \emph{{Cage-Net Fracton Models}},
  \href{http://dx.doi.org/10.1103/PhysRevX.9.021010}{\emph{Phys. Rev.} {\bf X9}
  021010 (2019)}, [\href{https://arxiv.org/abs/1806.04687}{{\tt
  arXiv:1806.04687}}].

\bibitem{BulmashMaissamBarkeshli2019taq1905.05771}
D.~Bulmash and M.~Barkeshli, \emph{{Gauging fractons: immobile non-Abelian
  quasiparticles, fractals, and position-dependent degeneracies}},
  \href{https://arxiv.org/abs/1905.05771}{{\tt arXiv:1905.05771}}.

\bibitem{PremWilliamson2019etl1905.06309}
A.~Prem and D.~J. Williamson, \emph{{Gauging permutation symmetries as a route
  to non-Abelian fractons}},  \href{https://arxiv.org/abs/1905.06309}{{\tt
  arXiv:1905.06309}}.

\bibitem{DijkgraafWitten1989pz1990}
R.~Dijkgraaf and E.~Witten, \emph{{Topological Gauge Theories and Group
  Cohomology}},
  \href{http://dx.doi.org/10.1007/BF02096988}{\emph{Commun.Math.Phys.} {\bf
  129} 393 (1990)}.

\bibitem{1602.05951WWY}
J.~Wang, X.-G. Wen and S.-T. Yau, \emph{{Quantum Statistics and Spacetime
  Surgery}},  \href{https://arxiv.org/abs/1602.05951}{{\tt arXiv:1602.05951}}.

\bibitem{Putrov2016qdo1612.09298}
P.~Putrov, J.~Wang and S.-T. Yau, \emph{{Braiding Statistics and Link
  Invariants of Bosonic/Fermionic Topological Quantum Matter in 2+1 and 3+1
  dimensions}}, \href{http://dx.doi.org/10.1016/j.aop.2017.06.019}{\emph{Annals
  Phys.} {\bf 384} 254--287 (2017)},
  [\href{https://arxiv.org/abs/1612.09298}{{\tt arXiv:1612.09298}}].

\bibitem{Wang2018edf1801.05416}
J.~Wang, K.~Ohmori, P.~Putrov, Y.~Zheng, Z.~Wan, M.~Guo et~al.,
  \emph{{Tunneling Topological Vacua via Extended Operators: (Spin-)TQFT
  Spectra and Boundary Deconfinement in Various Dimensions}},
  \href{http://dx.doi.org/10.1093/ptep/pty051}{\emph{PTEP} {\bf 2018} 053A01
  (2018)}, [\href{https://arxiv.org/abs/1801.05416}{{\tt arXiv:1801.05416}}].

\bibitem{Wang1901.11537WWY}
J.~Wang, X.-G. Wen and S.-T. Yau, \emph{{Quantum Statistics and Spacetime
  Topology: Quantum Surgery Formulas}},
  \href{http://dx.doi.org/10.1016/j.aop.2019.06.002}{\emph{Annals Phys.} {\bf
  409} 167904 (2019)}, [\href{https://arxiv.org/abs/1901.11537}{{\tt
  arXiv:1901.11537}}].

\bibitem{Ginzburg1950srGinzburgLandau}
V.~L. Ginzburg and L.~D. Landau, \emph{{On the Theory of superconductivity}},
  {\emph{Zh. Eksp. Teor. Fiz.} {\bf 20} 1064--1082 (1950)}.

\bibitem{Fisher-Lee}
M.~P.~A. Fisher and D.~H. Lee, \emph{Correspondence between two-dimensional
  bosons and a bulk superconductor in a magnetic field},
  \href{http://dx.doi.org/10.1103/PhysRevB.39.2756}{\emph{Phys. Rev. B} {\bf
  39} 2756--2759 (1989 Feb)}.

\bibitem{Dasgupta-Halperin}
C.~Dasgupta and B.~I. Halperin, \emph{Phase transition in a lattice model of
  superconductivity},
  \href{http://dx.doi.org/10.1103/PhysRevLett.47.1556}{\emph{Phys. Rev. Lett.}
  {\bf 47} 1556--1560 (1981 Nov)}.

\bibitem{Nelson}
D.~R. Nelson, \emph{Vortex entanglement in high-${T}_{c}$ superconductors},
  \href{http://dx.doi.org/10.1103/PhysRevLett.60.1973}{\emph{Phys. Rev. Lett.}
  {\bf 60} 1973--1976 (1988 May)}.

\bibitem{Yuan2019gehPengYe1911.02876}
J.-K. Yuan, S.~Chen and P.~Ye, \emph{{Fractonic superfluids, topological
  vortices, and quantum fluctuations}},
  \href{https://arxiv.org/abs/1911.02876}{{\tt arXiv:1911.02876}}.

\bibitem{PhysRev96191YM1954}
C.~N. Yang and R.~L. Mills, \emph{{Conservation of Isotopic Spin and Isotopic
  Gauge Invariance}},
  \href{http://dx.doi.org/10.1103/PhysRev.96.191}{\emph{Phys. Rev.} {\bf 96}
  191--195 (1954 Oct)}.

\bibitem{Gu2015lfa1503.01768}
Z.-C. Gu, J.~C. Wang and X.-G. Wen, \emph{{Multi-kink topological terms and
  charge-binding domain-wall condensation induced symmetry-protected
  topological states: Beyond Chern-Simons/BF theory}},
  \href{http://dx.doi.org/10.1103/PhysRevB.93.115136}{\emph{Phys. Rev.} {\bf
  B93} 115136 (2016)}, [\href{https://arxiv.org/abs/1503.01768}{{\tt
  arXiv:1503.01768}}].

\bibitem{YeGu2015eba1508.05689}
P.~Ye and Z.-C. Gu, \emph{{Topological quantum field theory of
  three-dimensional bosonic Abelian-symmetry-protected topological phases}},
  \href{http://dx.doi.org/10.1103/PhysRevB.93.205157}{\emph{Phys. Rev.} {\bf
  B93} 205157 (2016)}, [\href{https://arxiv.org/abs/1508.05689}{{\tt
  arXiv:1508.05689}}].

\bibitem{1959PhRvLHubbard}
J.~{Hubbard}, \emph{{Calculation of Partition Functions}},
  \href{http://dx.doi.org/10.1103/PhysRevLett.3.77}{\emph{"Phys. Rev. Lett."}
  {\bf 3} 77--78 (1959 Jul)}.

\bibitem{Ye2014oua1410.2594PengYe}
P.~Ye and Z.-C. Gu, \emph{{Vortex-Line Condensation in Three Dimensions: A
  Physical Mechanism for Bosonic Topological Insulators}},
  \href{http://dx.doi.org/10.1103/PhysRevX.5.021029}{\emph{Phys. Rev.} {\bf X5}
  021029 (2015)}, [\href{https://arxiv.org/abs/1410.2594}{{\tt
  arXiv:1410.2594}}].

\bibitem{evans10}
L.~C. Evans, \emph{Partial differential equations}.
\newblock American Mathematical Society, Providence, R.I., 2010.

\bibitem{2018PRLPretkoLeo1711.11044}
M.~{Pretko} and L.~{Radzihovsky}, \emph{{Fracton-Elasticity Duality}},
  \href{http://dx.doi.org/10.1103/PhysRevLett.120.195301}{\emph{Phys. Rev.
  Lett.} {\bf 120} 195301 (2018 May)},
  [\href{https://arxiv.org/abs/1711.11044}{{\tt arXiv:1711.11044}}].

\bibitem{2018PRLPretkoLeo1808.05616}
M.~{Pretko} and L.~{Radzihovsky}, \emph{{Symmetry-Enriched Fracton Phases from
  Supersolid Duality}},
  \href{http://dx.doi.org/10.1103/PhysRevLett.121.235301}{\emph{Phys. Rev.
  Lett.} {\bf 121} 235301 (2018 Dec)},
  [\href{https://arxiv.org/abs/1808.05616}{{\tt arXiv:1808.05616}}].

\bibitem{2019LeoHermelearXiv1905.06951}
L.~{Radzihovsky} and M.~{Hermele}, \emph{{Fractons from vector gauge theory}},
  {\emph{arXiv e-prints} arXiv:1905.06951 (2019 May)},
  [\href{https://arxiv.org/abs/1905.06951}{{\tt arXiv:1905.06951}}].

\bibitem{2019PretkoarXivCrystal-to-Fracton-1907.12577}
M.~{Pretko}, Z.~{Zhai} and L.~{Radzihovsky}, \emph{{Crystal-to-Fracton Tensor
  Gauge Theory Dualities}}, {\emph{arXiv e-prints} arXiv:1907.12577 (2019
  Jul)}, [\href{https://arxiv.org/abs/1907.12577}{{\tt arXiv:1907.12577}}].

\bibitem{Gromov2019waaPiotr1908.06984}
A.~Gromov and P.~Surowka, \emph{{On duality between Cosserat elasticity and
  fractons}},  \href{https://arxiv.org/abs/1908.06984}{{\tt arXiv:1908.06984}}.

\bibitem{YouPretko2019cvs1908.08540plaquette-melting-transitions}
Y.~You, Z.~Bi and M.~Pretko, \emph{{Emergent fractons and algebraic quantum
  liquid from plaquette melting transitions}},
  \href{https://arxiv.org/abs/1908.08540}{{\tt arXiv:1908.08540}}.

\bibitem{Vijay2017cti1706.07070}
S.~Vijay and L.~Fu, \emph{{A Generalization of Non-Abelian Anyons in Three
  Dimensions}},  \href{https://arxiv.org/abs/1706.07070}{{\tt
  arXiv:1706.07070}}.

\bibitem{SongPremHuang2018gbb1805.06899}
H.~Song, A.~Prem, S.-J. Huang and M.~A. Martin-Delgado, \emph{{Twisted Fracton
  Models in Three Dimensions}},
  \href{http://dx.doi.org/10.1103/PhysRevB.99.155118}{\emph{Phys. Rev.} {\bf
  B99} 155118 (2019)}, [\href{https://arxiv.org/abs/1805.06899}{{\tt
  arXiv:1805.06899}}].

\bibitem{2001.01722PretkoChenYou}
M.~Pretko, X.~Chen and Y.~You, \emph{{Fracton Phases of Matter}},
  \href{http://dx.doi.org/10.1142/S0217751X20300033}{\emph{Int. J. Mod. Phys.
  A} {\bf 35} 2030003 (2020)}, [\href{https://arxiv.org/abs/2001.01722}{{\tt
  arXiv:2001.01722}}].

\bibitem{Wan1211.3695}
Y.~{Hu}, Y.~{Wan} and Y.-S. {Wu}, \emph{{Twisted quantum double model of
  topological phases in two dimensions}},
  \href{http://dx.doi.org/10.1103/PhysRevB.87.125114}{\emph{Phys. Rev. B} {\bf
  87} 125114 (2013 Mar.)}, [\href{https://arxiv.org/abs/1211.3695}{{\tt
  arXiv:1211.3695}}].

\bibitem{Wan2014woaWWH1409.3216}
Y.~Wan, J.~C. Wang and H.~He, \emph{{Twisted Gauge Theory Model of Topological
  Phases in Three Dimensions}},
  \href{http://dx.doi.org/10.1103/PhysRevB.92.045101}{\emph{Phys. Rev.} {\bf
  B92} 045101 (2015)}, [\href{https://arxiv.org/abs/1409.3216}{{\tt
  arXiv:1409.3216}}].

\bibitem{Song2018gbb1805.06899}
H.~Song, A.~Prem, S.-J. Huang and M.~A. Martin-Delgado, \emph{{Twisted Fracton
  Models in Three Dimensions}},
  \href{http://dx.doi.org/10.1103/PhysRevB.99.155118}{\emph{Phys. Rev.} {\bf
  B99} 155118 (2019)}, [\href{https://arxiv.org/abs/1805.06899}{{\tt
  arXiv:1805.06899}}].

\bibitem{JWangCMSA2019-QM}
J.~C. {Wang}, \emph{{Higher-Rank Tensor Non-Abelian Gauge Field Theory of
  Fracton and Embeddon}}, {\emph{Quantum Matter workshop, Quantum Matter in
  Mathematics and Physics at Harvard CMSA, December 3rd, 2019
  \href{https://www.youtube.com/watch?v=77vkcOrvW8k}{https://www.youtube.com/watch?v=77vkcOrvW8k}}
  (2019)}.

\end{thebibliography}\endgroup

\end{document}